\documentclass[usegraphicx,usenatbib,useAMS]{mn2e}
  
\usepackage{epsfig}

\newcommand{\lesssim}{\la}

\title[Large-scale non-Gaussian mass function and halo bias: tests on 
N-body simulations]{Large-scale non-Gaussian mass function and halo bias: 
tests on N-body simulations}

\author[Grossi et al.]{M. Grossi$^1$, 
L. Verde $^{2,3}$,  C. Carbone$^3$, K. Dolag$^1$, E. Branchini$^4$, F. Iannuzzi$^{1,5}$, \newauthor
S. Matarrese$^6$, L. Moscardini$^{5,7}$\\
$^{1}$Max-Planck Institut fuer Astrophysik, 
Karl-Schwarzschild Strasse 1, D-85748 Garching, Germany \\
(margot,dolag,iannuzzi@mpa-garching.mpg.de)\\
$^{2}$ICREA\\
$^3$Institute of Space Sciences (CSIC-IEEC),  UAB, Barcelona 08193, Spain (verde,carbone@ieec.uab.es)\\
$^4$Dipartimento di Fisica, Universit\`a di Roma TRE, 
via della Vasca Navale 84, I-00146, Roma, Italy (branchin@fis.uniroma3.it) \\
$^5$Dipartimento di Astronomia, Universit\`a di Bologna, 
via Ranzani 1, I-40127 Bologna, Italy (lauro.moscardini@unibo.it)\\
$^6$Dipartimento di Fisica ''G. Galilei", Universit\`a 
degli Studi di Padova and INFN Sezione di Padova, 
via Marzolo 8, I-35131, Padova, \\
Italy (sabino.matarrese@pd.infn.it)\\
$^7$INFN, Sezione di Bologna, viale Berti Pichat 
6/2, I-40127 Bologna, Italy}

\def\gs{\mathrel{\raise1.16pt\hbox{$>$}\kern-7.0pt 
\lower3.06pt\hbox{{$\scriptstyle \sim$}}}}         
\def\ls{\mathrel{\raise1.16pt\hbox{$<$}\kern-7.0pt 
\lower3.06pt\hbox{{$\scriptstyle \sim$}}}}         

\begin{document}
\maketitle

\begin{abstract}
The description of the abundance and clustering of halos for non-Gaussian 
initial conditions has recently received renewed interest, motivated by 
the forthcoming large galaxy and cluster surveys, which can potentially 
yield constraints of order unity on the non-Gaussianity parameter $f_{NL}$. 
We present tests on N-body simulations of analytical formulae describing 
the halo abundance and clustering for non-Gaussian initial conditions.
We calibrate the analytic non-Gaussian mass function  of \citet{MVJ00} and 
\citet{Loverdeetal07} and the analytic description of clustering of halos 
for non-Gaussian initial conditions on N-body simulations. 
We find excellent agreement between the simulations and the analytic  predictions if we make the corrections $\delta_c \longrightarrow \delta_c\times \sqrt{q}$ and $\delta_c\longrightarrow \delta_c \times q$ where $q\simeq 0.75$,  in the density threshold for gravitational collapse and in the
non-Gaussian fractional correction to the halo bias, respectively.
We discuss the implications of this correction on present and forecasted 
primordial non-Gaussianity constraints. We confirm that the non-Gaussian halo 
bias offers a robust and highly competitive test of primordial non-Gaussianity.
\end{abstract}

\section{Introduction} 
Constraining primordial non-Gaussianity offers a powerful test of the 
generation mechanism of cosmological perturbations in the early universe. 
While standard single-field models of slow-roll inflation lead to small 
departures from Gaussianity, non-standard scenarios allow for a larger 
level of non-Gaussianity (\citet{BKMR04} and references therein). 
The standard observables to constrain non-Gaussianity are the Cosmic 
Microwave Background (CMB) and the Large Scale Structure (LSS) of the Universe. 
A powerful technique is based on the abundance \citep{MVJ00, VJKM01, 
Loverdeetal07, Soda99, RB00, RGS00} and clustering \citep{GW86, MLB86, LMV88} 
of rare events, such as  dark matter density peaks, as they trace the 
tail of the underlying matter distribution. 
Theoretical predictions on various observational aspects of non-Gaussianity have been  extensively tested against N-body simulations,
  leading to different and sometimes conflicting results \citep{KNS07,Grossietal07,DDHS07,Desjaques, Pillepich}. 

\citet{DDHS07} and \citet{MV08} showed that primordial non-Gaussianity 
affects the clustering of dark matter halos inducing a scale-dependent bias 
on large scales. Not only this effect has been already exploited to place 
stringent constraints on non-Gaussianity \citep{slosaretal08, 
AfshordiTolley08}, but also it is particularly promising for constraining 
non-Gaussianity from  future surveys, which will provide a large sample of 
galaxy clusters over a volume comparable to the horizon size 
(e.g., DES, PanSTARRS, BOSS, LSST, ADEPT, EUCLID) 
\citep{DDHS07,CVM08, AfshordiTolley08,seljak08}. 
\citet{bartolofnl05} showed that even for small primordial non-Gaussianities, 
the evolution of perturbations on super-Hubble scales yields extra 
contributions. The amplitude of these contributions is comparable to the forecasted errors of some planned surveys, opening up the possibility of 
measuring them.

In light of this, it is important to use N-body experiments to test
  the validity of theoretical predictions for halo-bias in non-Gaussian
  framework. Indeed, all proposed analytic biasing expressions have been
  derived in the extended Press-Schechter framework which assumes spherical
  collapse dynamics, sharp k-space filtering and Gaussian initial conditions.  
The validity of the extrapolation of the 
extended Press-Schechter approach to the non-Gaussian case can be tested 
independently by considering also the halo mass function. It is thus also important 
to test and calibrate on N-body simulations the predictions of the 
non-Gaussian halo mass function \citep{KNS07,Grossietal07,DDHS07} and of 
the non-Gaussian halo bias simultaneously. This is what we set out to do here.
  
In this paper we start by reviewing the analytic predictions for the Gaussian 
and non-Gaussian halo abundance and clustering (\S 2). In \S 3 we describe 
the numerical simulations with Gaussian and non-Gaussian initial conditions.  
In \S 4 we present the test for the non-Gaussian mass function.  
In \S 5 and 6 we test the analytic predictions of Gaussian and non-Gaussian 
large scale bias against N-body simulations. 
In \S 7 we compare our results with the literature. 
Finally, we conclude in \S 8.
  
\section{Formulation of the non-Gaussian halo abundance and clustering}    
Deviations from Gaussian initial conditions are commonly parameterized in 
terms of the dimensionless $f_{\rm NL}$ parameter 
\citep{SalopekBond90, Ganguietal94,VWHK00, KS01}:
\begin{equation}
\Phi=\phi+f_{\rm NL}(\phi^2-\langle\phi^2\rangle)\,,
\label{eq:local}
\end{equation} 
where $\Phi$ denotes the gravitational potential and $\phi$ is a 
Gaussian random field.
As noted by e.g., \cite{Loverdeetal07}, \cite{AfshordiTolley08} and 
\cite{Pillepich}, different authors use different conventions. 
Here $\Phi$ denotes Bardeen's gauge-invariant potential which, on sub-Hubble 
scales, reduces to the usual Newtonian peculiar gravitational potential 
but with a negative sign. In addition, there are two conventions for 
normalizing Eq.~(\ref{eq:local}): the LSS and the CMB one. 
In the LSS convention $\Phi$ is linearly extrapolated at $z = 0$. 
In the present paper we use this convention.  In the CMB convention $\Phi$ 
is instead primordial: thus $f_{\rm NL} = 
[g(z = \infty )/g(0)]f_{\rm NL}^{CMB}\sim 1.3 f^{CMB}_{NL}$, where 
$g(z)$ denotes the linear growth suppression factor in non Einstein-de-Sitter 
Universes.

\subsection{Formulation of the non-Gaussian mass function: 
Extended Press-Schechter approach}

In the Press-Schechter framework, one considers the density contrast field 
evaluated at some early time, far before any scale of interest has approached 
the nonlinear regime, but extrapolated to 
the present day using linear perturbation theory. 
Then one considers the height of the critical density threshold as a function 
of time. In that way, the collapse of a halo at redshift $z \ne 0$ corresponds 
to the $z = 0$ density fluctuation crossing a barrier of height 
$\delta_c (z) = \Delta_cD(z=0)/D(z)$, where $\Delta_c \sim \delta_c(z=0)$ 
(this is an equality only in an Einstein de Sitter Universe); 
we use $D(z=0)=1$, $D(z)=g(z)/g(0)(1+z)^{-1}$.
We should recall here that, even in linear theory, the normalized skewness 
of the density field, $S_3\equiv \langle \delta^3 \rangle/\langle \delta^2 
\rangle^2$, depends on redshift $\propto 1/D(z)$, however in the 
Press-Schechter framework one should use the linear $S_3(z=0)$, 
in what follows 
$S_3\equiv S_3(z=0)$.
Note also that in general the skewness can be written as 
$S_3 \equiv f_{\rm NL}S_3^{(1)}$, where $S_3^{(1)}$ denotes the skewness in 
units of $f_{\rm NL}$, care must be exercised in the  interpretation of 
$f_{\rm NL}$: if $S_3^{(1)}$ is that of the  density field linearly 
extrapolated at $z=0$, $f_{NL}$ must be the LSS one and {\it not} the CMB one.

Generalization of the mass function to non-Gaussian initial conditions within 
the Press-Shechter formalism has been presented in \cite{MVJ00, Loverdeetal07}.
Both references start by computing an expression for the non-Gaussian 
probability density function of the smoothed dark matter density field, 
then obtain the level excursion  probability. In the Press-Shechter approach 
the mass derivative of the level excursion  probability is the key ingredient 
to obtain the mass function expression and is the term that gets modified 
in the presence of primordial non-Gaussianity. In this derivation, 
several approximations are made. Both approaches assume that deviations from 
Gaussianity are small. 

\cite{MVJ00} use  first the saddle-point 
approximation to compute the level excursion probability and then truncate 
the resulting expression at the skewness. 
They obtain \footnote{We correct here a typo in Eq.~(68) of \cite{MVJ00}, 
where  $d\ln \sigma_M$ should be $d\sigma_M$.}:
\begin{eqnarray}
n(M,z)&=&2\frac{3H_0^2\Omega_{m,0}}{8\pi GM^2}\frac{1}{\sqrt{2\pi}\sigma_M} 
\exp\left[-\frac{\delta_*^2}{2\sigma_M^2}\right] \times \\
&&\left| \frac{1}{2}\frac{\delta_c^2}{3\sqrt{1-S_{3,M}\delta_c/3}} 
\frac{dS_{3,M}}{d\ln M} +\frac{\delta_*}{\sigma_M}
\frac{d \sigma_M}{d\ln M}\right|  \nonumber 
\label{eq: massfnMVJ}
\end{eqnarray}
where $\sigma_M$ denotes the {\it rms} value of the density field, the 
subscript $M$ denotes that the density field has been smoothed on a scale 
$R(M)$ corresponding to $R(M)=[M 3/(4\bar{\rho}_M)]^{1/3}$, and 
$\delta_*=\delta_c\sqrt{1-\delta_c S_{3,M}/3}$. 

\cite{Loverdeetal07} instead first approximate the probability density 
function using the Edgeworth expansion, then perform the integral of the 
level excursion probability exactly on the first few terms of the expansion. 
They obtain:
\begin{eqnarray}
n(M,z)&=&2\frac{3H_0^2\Omega_{m,0}}{8\pi GM^2}\frac{1}{\sqrt{2\pi}\sigma_M} 
\exp\left[-\frac{\delta_c^2}{2\sigma_M^2}\right] \times \\
& & \left[  \frac{d \ln \sigma_M}{dM}\left(\frac{\delta_c}{\sigma_M}+\right. 
\frac{S_{3,M}\sigma_M}{6} \left( \frac{\delta_c^4}{\sigma_M^4}
-2\frac{\delta_c^2}{\sigma_M^2}-1\right) \right)  \nonumber  \\
&& \left. +\frac{1}{6}\frac{dS_{3,M}}{dM}
\sigma_M\left(\frac{\delta_c^2}{\sigma_M^2}-1\right)  \right] \nonumber
\label{eq:massfnloverde}
\end{eqnarray}
Note that in the limit of small non-Gaussianity and rare events, the ratio of 
the non-Gaussian mass function to the Gaussian one for both expressions 
reduces to:
\begin{equation}
{\cal R}_{NG}\equiv \frac{n(M,z|f_{\rm NL})}{n(M,z|f_{\rm NL}=0)}
\longrightarrow 1+S_{3,M}\frac{\delta_c^3}{6\sigma_M^2}\,.
\end{equation}
It is important to bear in mind that  in Eqs. (\ref{eq: massfnMVJ})-(\ref{eq:massfnloverde}), 
the redshift dependence is enclosed only in $\delta_c$ (and not in $S_3$). 
In the spirit of the ``CMB'' convention instead, where the gravitational 
potential is normalized deep in the matter era, one should  make sure that all
the relevant quantities are correctly extrapolated linearly at $z=0$, 
keeping in mind that the gravitational potential slowly evolves in a 
non Einstein de Sitter Universe.

The major limitations in both derivations are the assumption of spherical 
collapse and the sharp $k$-space filtering. In addition, the excursion set 
improvement on the interpretation of the original Press-Shechter swindle, 
suggests that this derivation relies on the random-phase hypothesis 
\cite{Sheth98}, which is clearly not satisfied for non-Gaussian 
initial conditions even for sharp $k$-space filtering.

\cite{VJKM01} and \cite{Loverdeetal07} addressed this issue by using the 
analytical approach to compute the fractional non-Gaussian {\it correction} 
to the Gaussian mass function ${\cal R}_{NG}$, and used the \cite{ST} 
mass function to model the Gaussian mass function. 
This approach is potentially promising, but needs to be calibrated on numerical experiments.

In particular, one may argue that the same correction that in
the Gaussian case modifies the collapse threshold and thus the form
of the mass function from \citet{PS} to \citet{SMT01} and \citet{ST02}, may
apply to the non-Gaussian correction. In the Gaussian case this is
usually referred to as the correction due to ellipsoidal collapse
\citep{LeeShandarin}.  While this interpretation has recently been
disputed (see  e.g., \citet{Robertson08}),  we will maintain the same
nomenclature here.  For rare events, high peaks ($\delta_c/\sigma_M
\gg1$) and small $f_{\rm NL}$,  this is equivalent to lower
$\delta_c$ by a factor $\sqrt{q}$ with $q=0.75$.

In summary we propose that the non-Gaussian mass function $n(M,z,f_{NL})$ 
should be re-written in terms of the Gaussian one $n^{sim}_G(M,z)$ --given by 
tested fits to simulations  e.g., \citet{ST,Reed, Warren,Jenkins}--, multiplied 
by a non-Gaussian correction factor:
\begin{equation}
n(M,z,f_{NL})=n^{sim}_G(M,z)\times {\cal R}_{NG}(M,z,f_{NL})
\label{eq:NGmassfn}
\end{equation}
where $R_{NG}(M,z,f_{NL})$ takes two different forms in the \cite{MVJ00} 
and \cite{Loverdeetal07} approximations.
For the \cite{MVJ00} case \footnote{We correct here a typo in Eq.~(3) of 
\cite{Grossietal07} where the exponential part was missing.}
\begin{eqnarray}
\label{eq:ratioMVJellips}
&&R_{NG}(M,z,f_{NL})=\exp\left[\delta_{ec}^3
\frac{S_{3,M}}{6 \sigma_M^2}\right] \times \\
& &\!\!\!\! \left| \frac{1}{6}
\frac{\delta_{ec}^2}{\sqrt{1-\delta_{ec}S_{3,M}/3}} 
\frac{dS_{3,M}}{d\ln \sigma_{M}} \right. %\nonumber \\
+\left. \frac{\delta_{ec}\sqrt{1-\delta_{ec} S_{3,M}/3}}{\delta_{ec}}\right|  
\nonumber 
\end{eqnarray}
and for the \cite{Loverdeetal07}  case:
\begin{eqnarray}
\label{eq:ratioLoVellips}
&&R_{NG}(M,z,f_{NL})=1+\frac{1}{6}\frac{\sigma_M^2}{\delta_{ec}} \times \\
&&\left[S_{3,M}\left(\frac{\delta_{ec}^4}{\sigma_M^4} 
-2\frac{\delta_{ec}^2}{\sigma_M^2}-1\right)+
\frac{dS_{3,M}}{d \ln \sigma_M}
\left(\frac{\delta_{ec}^2}{\sigma_M^2}-1\right)\right] \nonumber
\end{eqnarray}
where $\delta_{ec}$ denotes the critical density for ellipsoidal collapse, 
which for high peaks is $\delta_{ec}\sim \delta_c \sqrt{q}$ with $q=0.75$.

\subsection{Formulation of the non-Gaussian large scale halo bias}

For the case of ``local" primordial non-Gaussianity Eq.~(\ref{eq:local}),
the analytical expression for the large-scale non-Gaussian bias has been 
derived in five different ways, obtaining always basically the same result. 
\cite{DDHS07} considered the Laplacian of $\Phi$ in the vicinity of rare, 
high peaks, considering that the resulting $\nabla^2\Phi$ is proportional 
to the peaks overdensity; they also generalized to local non-Gaussianity the 
\cite{Kaiser84} argument of high-peaks bias in order to derive its 
non-Gaussian version. \cite{MV08} derived the halo bias formula in general 
non-Gaussian cases specified by an expression for the bispectrum.
\cite{slosaretal08} adopted the peak-background split approach \citep{CK89} 
for the local non-Gaussian case, showing that the resulting  expression 
relies on the universality of the mass function. 
\cite{AfshordiTolley08} instead interpreted non-Gaussianity as a 
modification of the critical density for collapse, in the framework of 
ellipsoidal collapse.   
Finally, \cite{Mcdonald08} used a renormalized perturbation theory approach 
to consider at the same time non-linear bias, second-order gravitational 
evolution and local form of non-Gaussianity.     
It is encouraging that these different approaches yield a consistent result 
for the correction to the Gaussian Lagrangian halo bias $b_L^G$:
\begin{equation}
\frac{\Delta b}{b_L^G}=2 f_{\rm NL}\delta_c(z)\alpha_M(k)
\label{eq:ngcorrection}
\end{equation}
where $\alpha_M(k)$ encloses the scale and halo mass dependence --see e.g., 
Eq.~(13) and Fig.~3 of \citet{MV08}--. 
Also in this case the density field is the one extrapolated linearly 
at $z=0$, and  $\alpha_M$ does not depend on redshift. 

Making the standard assumption that halos move coherently 
with the underlying dark matter, the Lagrangian bias is related to 
the Eulerian one as  $b=1+b_L$.

The approximations used to derive this equation are Press-Schechter \citep{PS} 
approach, linear bias, small non-Gaussianity, and in most cases spherical 
collapse and identification of peaks with halos. 
It is therefore important to test the validity of Eq.~(\ref{eq:ngcorrection}), 
with simulations and see if any correction factor needed is indeed due to 
account for non-spherical collapse.
Following the derivation of \cite{MV08} we recognize that the correction to 
the 2-point halo correlation function due to non-Gaussianity (their Eq.~(6)) 
is multiplied by $\nu^3/\sigma_M^3$ with $\nu=\delta_c/\sigma_M$. 
In this factor we recognize one Lagrangian Gaussian bias factor to the 
second power and an extra $\delta_c/\sigma_M^2$, which denominator was absorbed 
in the form factor.
Recall that, as discussed in \S 2.1,  for ``ellipsoidal collapse" and rare events, the Lagrangian Gaussian 
bias is corrected as $\nu/\sigma_M \longrightarrow q \nu/\sigma_M$ (see 
Eq.~(\ref{eq:gaussianbias}) below, for high $\nu$).
However, the remaining factor is also a Gaussian bias and it should also be 
corrected by the $q$-factor.

We conclude that the ``non-spherical collapse" modifies 
Eq.~(\ref{eq:ngcorrection}) to be:
\begin{equation}
\frac{\Delta b}{b_L^G} \simeq 2 f_{\rm NL}\delta_c(z)\alpha_M(k)q \,.
\label{eq:ngcorrection2}
\end{equation}
Note that \cite{AfshordiTolley08} arrived to a similar yet not identical 
expression when considering ellipsoidal collapse, i.e. they suggest that 
$\delta_c$ should be substituted by the critical density of \cite{SMT01}, 
which in our limit would correspond to use $\sqrt{q}$ rather than $q$ in 
Eq.~(\ref{eq:ngcorrection2}).

In \S \ref{sec:ngbias} we will show that Eq.~(\ref{eq:ngcorrection2}) 
correction fits well the simulations.
\begin{figure}
%Figure 1
\includegraphics[width=0.5\textwidth]{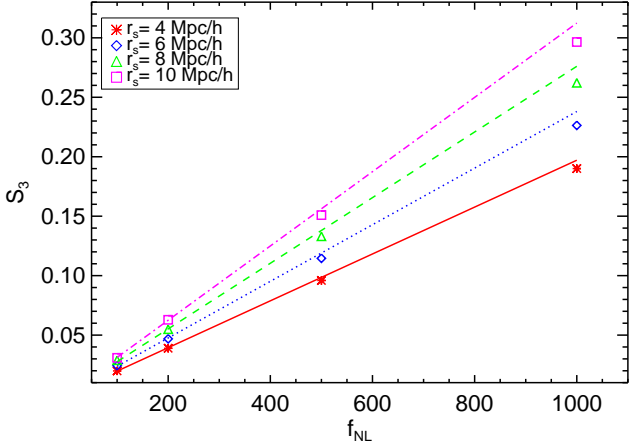}
 \caption{Skewness $S_3$ of the smoothed initial density field  for
$f_{\rm NL}=100,200,500,1000$. Symbols show the numerical results of the 
initial conditions code (averaged over 5 realizations) and are plotted 
against the analytical predictions for smoothing radii 
$r_s=4, 6, 8, 10 Mpc/h$ of a spherical top-hat filter.}
 \label{fig:s3}
 \end{figure}

\section{N-body simulations}

\begin{figure}
%Fig3 
\includegraphics[width=0.5\textwidth]{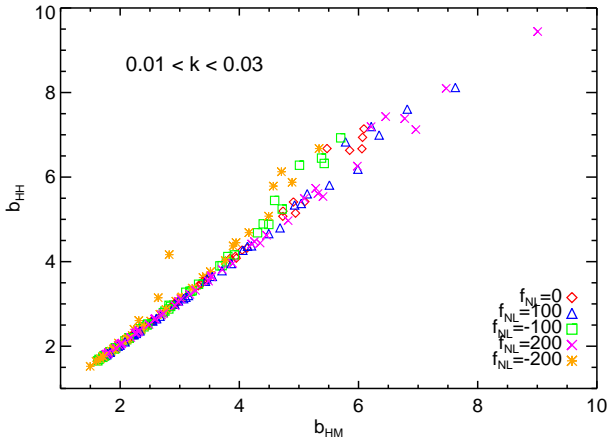}
\caption{ The bias of the halo power spectrum $b_{hh}$ compared to the bias of the cross (halo-matter) power spectrum  $b_{hm}$.  As expected, when the number density of halos is high 
there is good agreement between the two quantities. At low halo number densities the two quantities are affected differently by shot noise,  with $b_{hm}$ being the least affected.}
\label{fig:bhhbhm}
\end{figure}
%
%\begin{figure}
%%fig 2 
%\includegraphics[width=0.42\textwidth]{MassFuncGauss.ps}
% \caption{Multiplicity mass function for the Gaussian simulation computed 
% using a Friends-of-Friends halo finder. Points denote the simulations results at different redshift: $z=0,0.44,1.02,1.53,2.26$ and  $3.23$ (top to bottom). Solid (green) lines are the Sheth \& Tormen (1999)  formula,  dashed (red) lines are the Warren et al. (2006) one and dotted (blue) are the  Jenkins et al. (2001).} \label{fig:Gaumassfn}
%\end{figure}
 
The deviations from Gaussianity we are after become important on very large 
scales $k \lesssim 0.03$h/Mpc and for massive halos. Therefore, one needs to 
perform N-body simulations on very large boxes, yet with enough resolution 
to identify massive virialized structures at different redshifts.

Suitable initial conditions have been set up following the
method described in more detail in \cite{Grossi08} (see also
\cite{Grossietal07, viel09}).  
In brief, a random realization of a Gaussian
gravitational potential, $\Phi_{\rm L}$, normalized to be the one linearly 
extrapolated at $z=0$, is generated in Fourier space, then it
is inverse-Fourier transformed back to real space and added to the 
non-Gaussian term, $\Phi_{\rm NL}= f_{\rm{NL}} \left(\Phi_{\rm L}^2 - 
\langle\Phi_{\rm L}^2\rangle \right)$. The resulting field 
$\Phi_L+\Phi_{\rm NL}$ that is linear and at $z=0$, is transformed back in 
Fourier space.
We eventually modulate the power-law spectrum using the transfer function and 
compute the corresponding density field, which we then scale back to the 
initial conditions redshift ($z=60$).
The corresponding gravitational potential is then used to displace particles 
according to the Zel'dovich approximation. 
This method allows one to simulate non-Gaussian models having power spectra 
which are all consistent with that of the Gaussian case 
and was already used by \cite{viel09}.

In order to check the reliability of the initial conditions generation,
we have performed a specific test: using $256^3$ particles in a box of 
size 1000 Mpc$/h$, primordial density fields (extrapolated linearly at $z=0$) 
were generated and smoothed using spherical top-hat filters of different 
radii $r_s=4$, $6$, $8$, $12$ Mpc/$h$. The
smoothed skewness was then extracted from the fields and compared to
the analytical prediction for $f_{\rm NL}=100,200,500,1000$, as shown in 
Figure \ref{fig:s3}.

The set of simulations used in this work assumes the `concordance'
$\Lambda$CDM model. We fix the relevant parameters consistently with
those derived from the analysis of the WMAP 5-year data \citep{WMAP5}: 
$\Omega_{m0}=0.26$ for the matter density parameter,
$\Omega_{\Lambda0}=0.74$ for the $\Lambda$ contribution to the density
parameter, $h=0.72$ for the Hubble parameter (in units of $100$ km
s$^{-1}$ Mpc$^{-1}$). The initial power spectrum adopts the Cold Dark
Matter (CDM) transfer function suggested by 
\cite{EisensteinHu99}, has a spectral index $n=0.96$ and is normalized in 
such a way that $\sigma_{8}=0.8$. 
In all experiments, performed using the GADGET-2
numerical code \citep{GADGET}, switching off the hydrodynamical part,
we consider a box of (1200 Mpc/$h)^3$ with $960^{3}$ particles: the
corresponding particle mass is then $m\approx 1.4 \times 10^{11}
h^{-1} M_\odot$. The gravitational force has a Plummer-equivalent
softening length of $\epsilon_{l}=25 h^{-1}$ kpc. The runs produced
15 outputs from the initial redshift ($z=60$) to the present time.
The 5 simulations consider different amounts of primordial
non-Gaussianity, parametrized by the $f_{\rm NL}$ parameter: $f_{\rm
 NL}=0$ (i.e. the reference Gaussian case) and $f_{\rm NL}=\pm 100,
\pm 200$. 
The catalogues of dark matter haloes are extracted from the 
simulations using the standard friends-of-friends algorithm adopting a
linking length of $0.2$ times the mean interparticle distance; only
objects with at least $32$ particles are  considered.

We thus measure the halo bias in the simulations as
\begin{equation}
b_s(k,M,z)=b_{hm}\equiv\frac{P_{hm}(k,z,M)}{P_{mm}(k,z)} \;,
\end{equation}
where $P_{hm}(k,z,M)$ denotes the cross-power spectrum of dark matter 
with halos of mass $M$ at scale $k$, and for the simulation snapshot at 
redshift $z$. Similarly $P_{mm}(k,z)$ denotes the dark matter power spectrum. 
Here and hereafter the subscript $s$ denotes quantities measured from the 
simulation.

\begin{figure}
%fig 3
\includegraphics[width=0.45\textwidth]{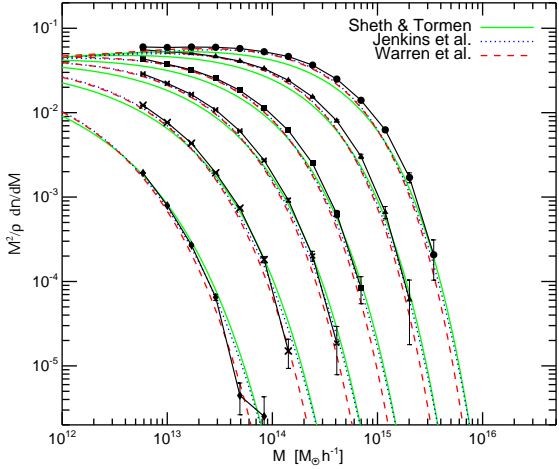}
 \caption{Multiplicity mass function for the Gaussian simulation computed 
 using a Friends-of-Friends halo finder. Points denote the simulations results at different redshift: $z=0,0.44,1.02,1.53,2.26$ and  $3.23$ (top to bottom). Solid (green) lines are the Sheth \& Tormen (1999)  formula,  dashed (red) lines are the Warren et al. (2006) one and dotted (blue) are the  Jenkins et al. (2001).} \label{fig:Gaumassfn}
\end{figure}

In principle, the quantity one is interested in would be the bias of the halo 
power spectrum $b_{hh}=\sqrt{P_{hh}/P_{mm}}$, but $b_{hm}$ is a less noisy 
quantity (the shot noise of the finite
number of halos is greatly suppressed in the estimate of
$P_{hm}(k)$). The quantity $b_{hm}$ is not guaranteed to be identical to 
$b_{hh}$ if bias has a stochastic component that does not correlate with the 
matter density field.
In Fig.~\ref{fig:bhhbhm} we show that this is not the case and that there is 
good agreement on large scales between $b_{hh}$ and $b_{hm}$, justifying using the less noisy $b_{hm}$ as an estimator for $b_{hh}$.

\begin{figure}
 \includegraphics[width=0.45\textwidth]{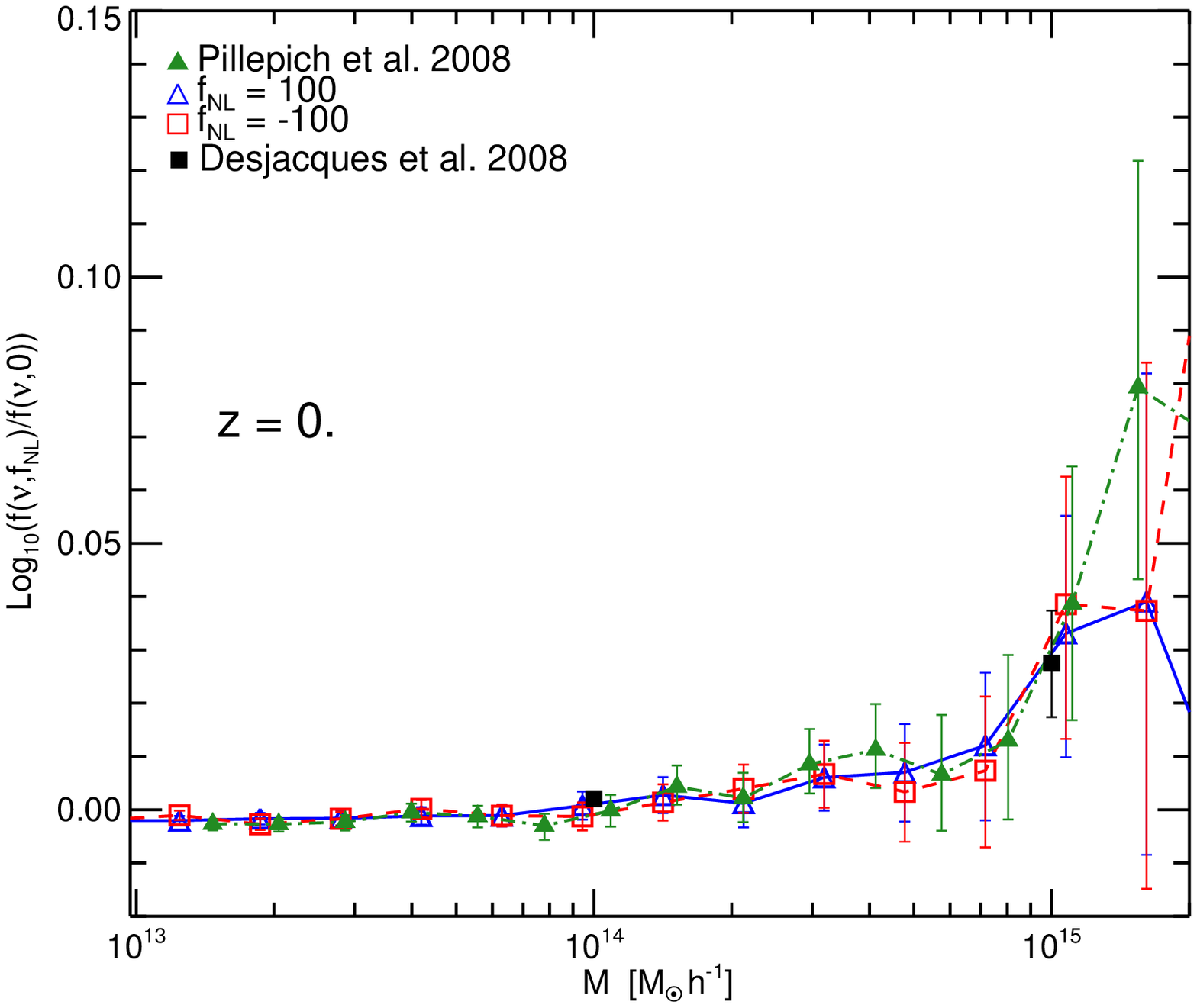}
\caption{Comparison between the halo mass function recovered in our
simulations with the work of \citet{Desjaques} and \citet{Pillepich} at 
$z=0$. We show the ratio between our non-Gaussian and Gaussian simulation 
with $f_{NL}=\pm 100$, few points we read out from Figure 1 of 
\citet{Desjaques}(black points) at the values of $\nu$ corresponding to 
$1\times 10^{13}$,$1\times 10^{14}$ and $1\times 10^{15}$ $M_{\odot}/h$ and 
the points from \citet{Pillepich}. We plot the reciprocal of the
results for $f_{NL}=-100$.}
\label{fig:comparisonsim1}
\end{figure}

\begin{figure*}
\begin{center}
\includegraphics[width=0.45\textwidth]{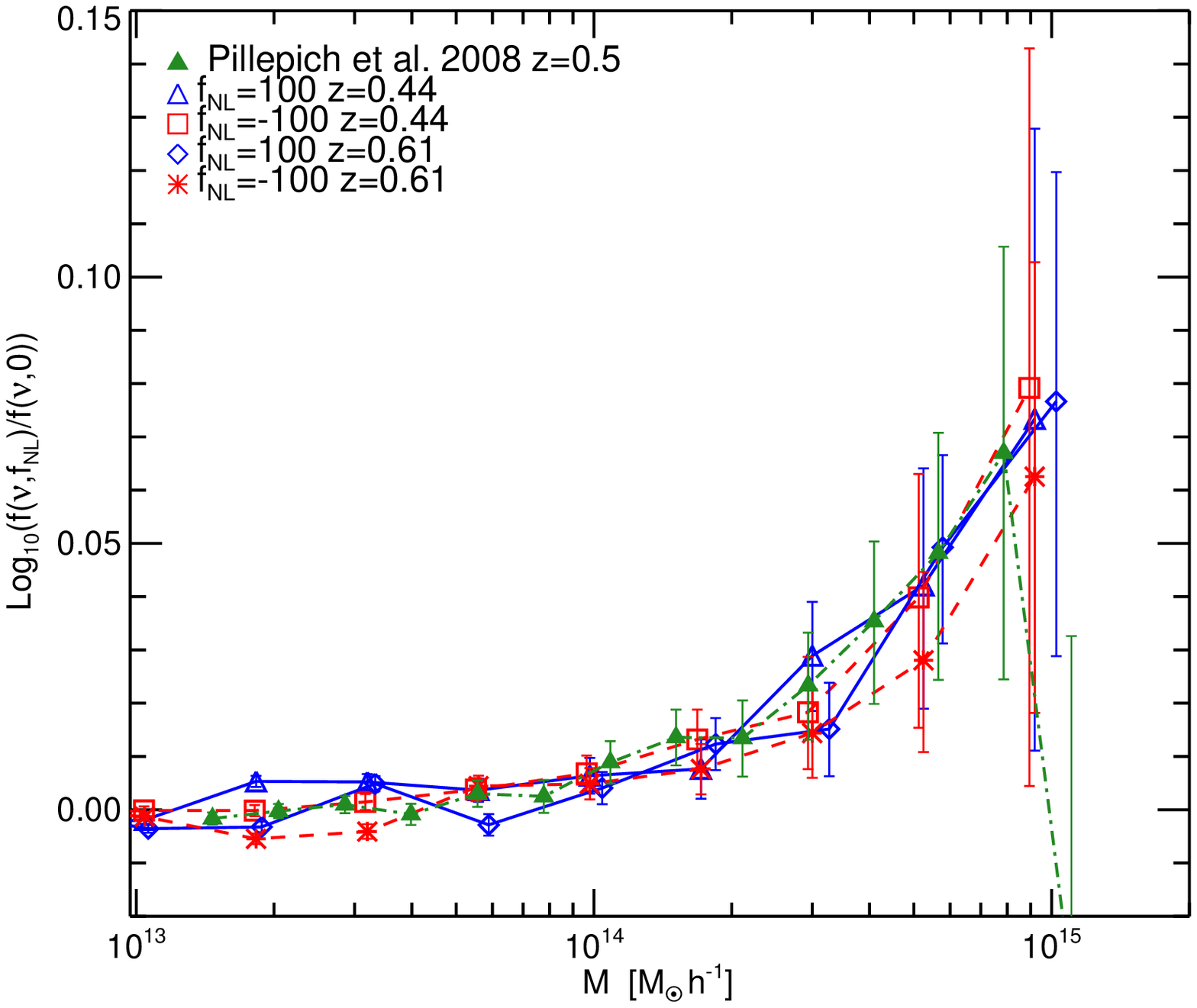}
\includegraphics[width=0.45\textwidth]{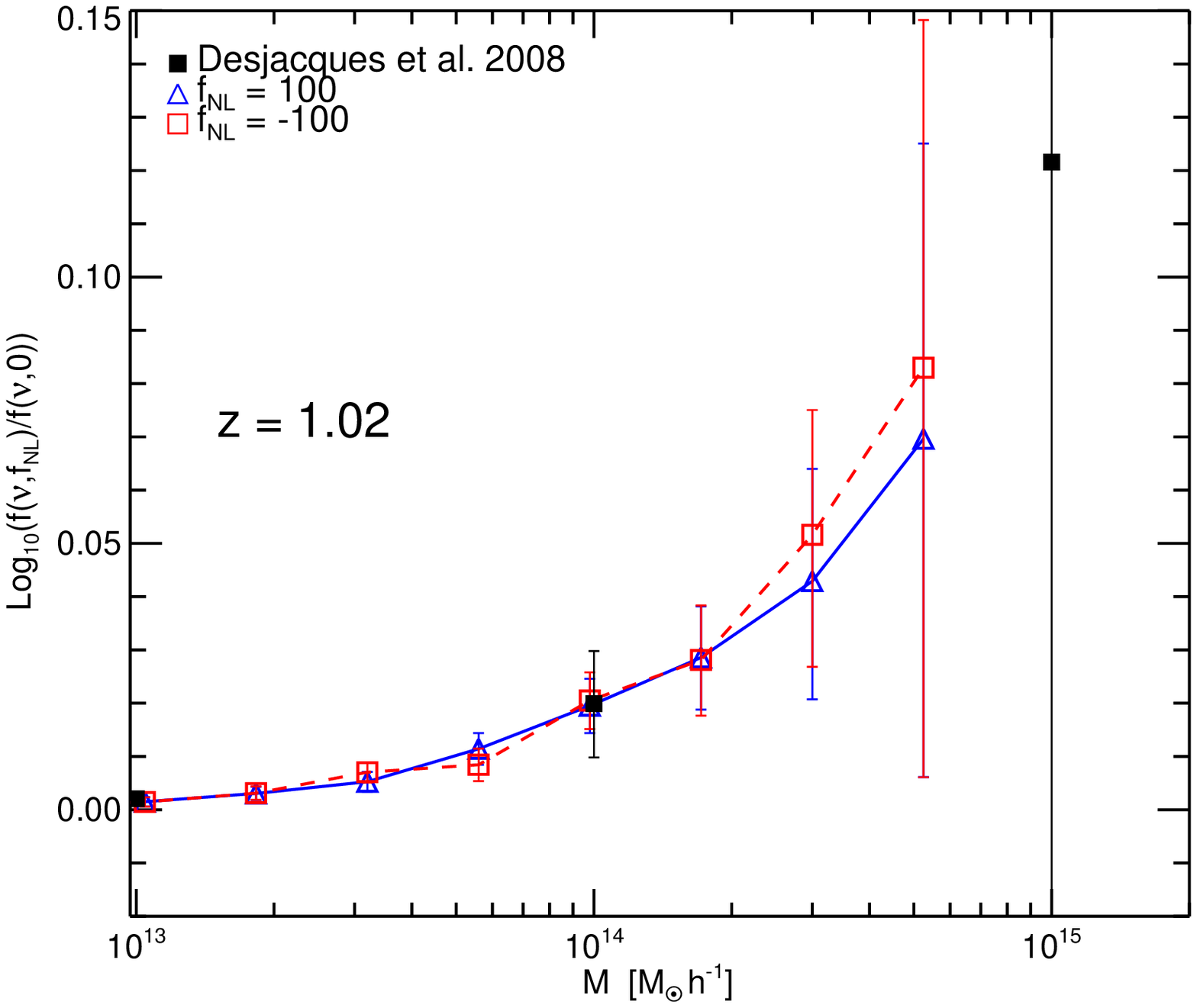}
\end{center}
\caption{Comparison between the halo mass function recovered in our
  simulations with the work of \citet{Desjaques} and \citet{Pillepich}.
  In the left panel we show the ratio between the non-Gaussian and Gaussian
  simulation at redshift $1$ for our
  simulations with $f_{NL}=\pm 100$, three points we read out from Figure 1
  of \citet{Desjaques}(black points) at the values of $\nu$ corresponding
  to $1\times 10^{13}$,$1\times 10^{14}$ and $1\times 10^{15}$ $M_{\odot}/h$.
  We plot the reciprocal of the results for $f_{NL}=-100$. In the right panel we show the data of
  Pillepich, Porciani, Hahn (2008) at $z=0.5$ and we compare them with our simulation results
  for the two closest available redshifts : $z=0.44$ and $z=0.61$ and with
  \citet{Desjaques}. All points are rescaled to $|f_{NL}=100|$ in our
  notation. The three independent simulations are in good agreement.}
\label{fig:comparisonsim2}
\end{figure*}

\subsection{Comparison with independent simulations}

In Fig.~\ref{fig:Gaumassfn} we show the mass function extracted from our 
Gaussian simulations at the following redshifts: $z=0.0,0.44,1.02,1.53,2.26$ 
and $3.23$. We also show three different theoretical predictions 
(also calibrated on N-body simulations): \citet{ST,Jenkins, Warren}, solid, 
dotted and dashed lines respectively. There is good  agreement even at high redshift.

Several groups recently presented N-body simulations, aiming at   quantifying  the effect of the non
Gaussian initial conditions on the halo mass function 
\citep{Desjaques,Pillepich}.
%,DDHS07}. 
All these results are obtained for similar 
cosmological parameters, so that we can compare estimates derived from all 
the simulations directly.
By comparing the results for the individual simulations at $z=1$, 
$z \sim 0.5$ and $z=0$ in Figures \ref{fig:comparisonsim1}, 
\ref{fig:comparisonsim2} we demonstrate
that these results are in agreement among the different groups, 
once the $f_{NL}$ values are suitably converted to the same convention.
Although all simulations use boxes of Giga parsec scales to explore the effect 
of non-Gaussian initial conditions at the high mass end, the statistical 
errors at the scale of massive clusters are still large. Therefore, we also 
report the reciprocal of the results obtained for negative $f_{NL}$ so that they appear in the positive part of the plot, to give an intuitive feeling of the 
noise within the individual simulations.

In Figure \ref{fig:comparisonsim1}, we show our simulation results for
$f_{NL}=100$ (blue triangles) and for $f_{NL}=-100$ (red squares) at
$z=0$ compared with data points from Figure 1 of \citet{Desjaques} 
(black points) at the
values of $\nu$ corresponding to $1\times 10^{13}$,$1\times 10^{14}$ and 
$1\times 10^{15}$ $M_{\odot}/h$ (as given in their figure caption). 
Note that, as \citet{Desjaques} use $f_{NL}=100$ in the CMB convention for 
their simulations, we scaled the points down accordingly by a factor 1.3 
to be comparable with our $f_{NL}=100$. We also show the results for 
\citet{Pillepich} (green points). Here we again apply the re-scaling as 
before, as their $f_{NL}$ of $82$ would correspond to a 
$f_{NL}$ of $\sim 106$ in the LSS notation.

In Figure \ref{fig:comparisonsim2}, the left panel shows the results for 
\citet{Pillepich} (green points) at $z=0.5$, and  our points
for the two closest available output times of our simulation
($z=0.44$ and $z=0.61$). 
 The right panel shows the comparison at $z=1$ between our points 
(blue triangles and red squares) and points from \citet{Desjaques} 
(black squares).

>From this comparison we conclude that there is remarkable agreement 
between the three independents simulations, highlighting the robustness of the simulations results. The differences visible
at some of the highest mass bins are not significant, given the large
error bars present.

\begin{figure*}
\begin{center}
\includegraphics[width=0.45\textwidth]{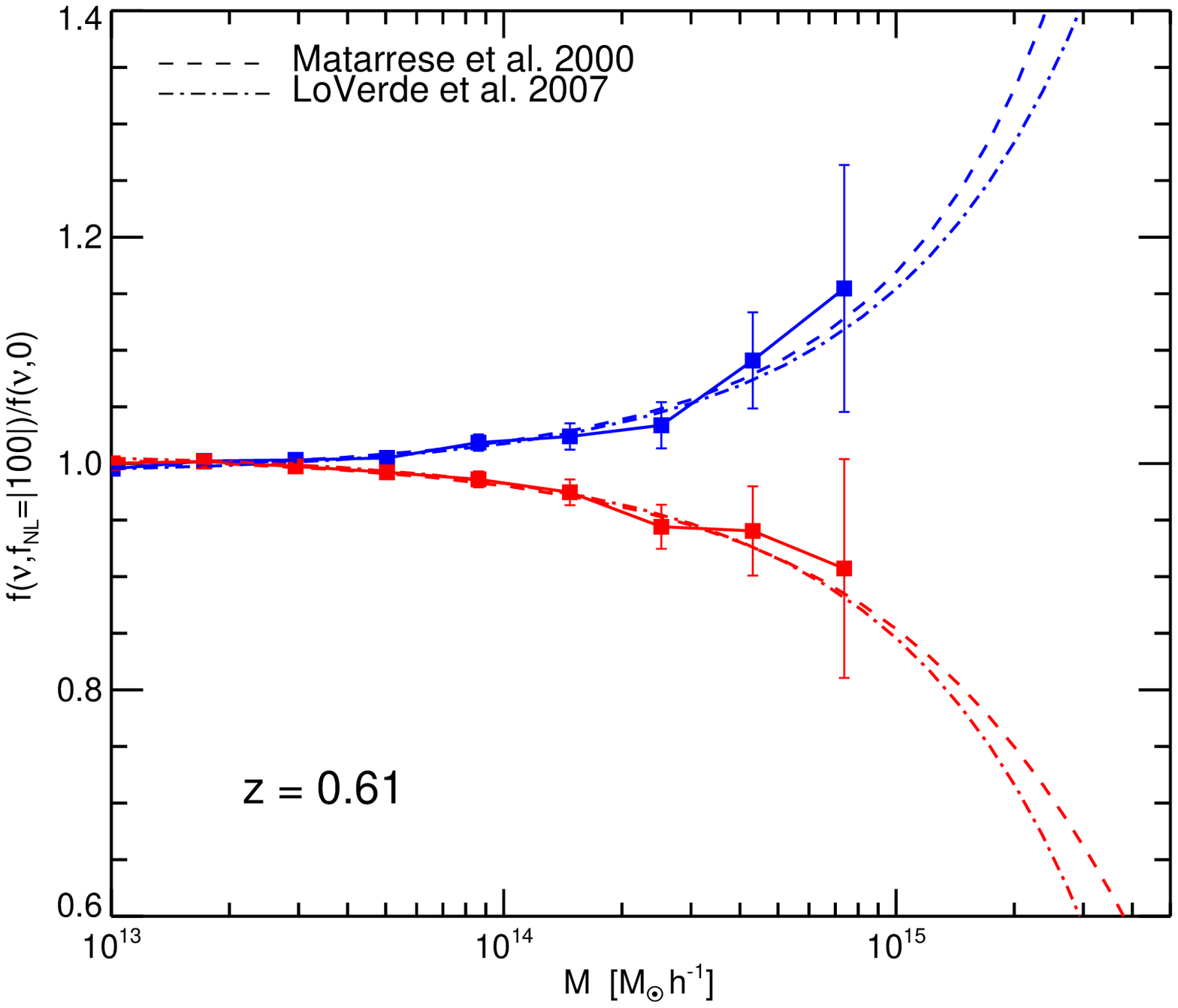}
\includegraphics[width=0.45\textwidth]{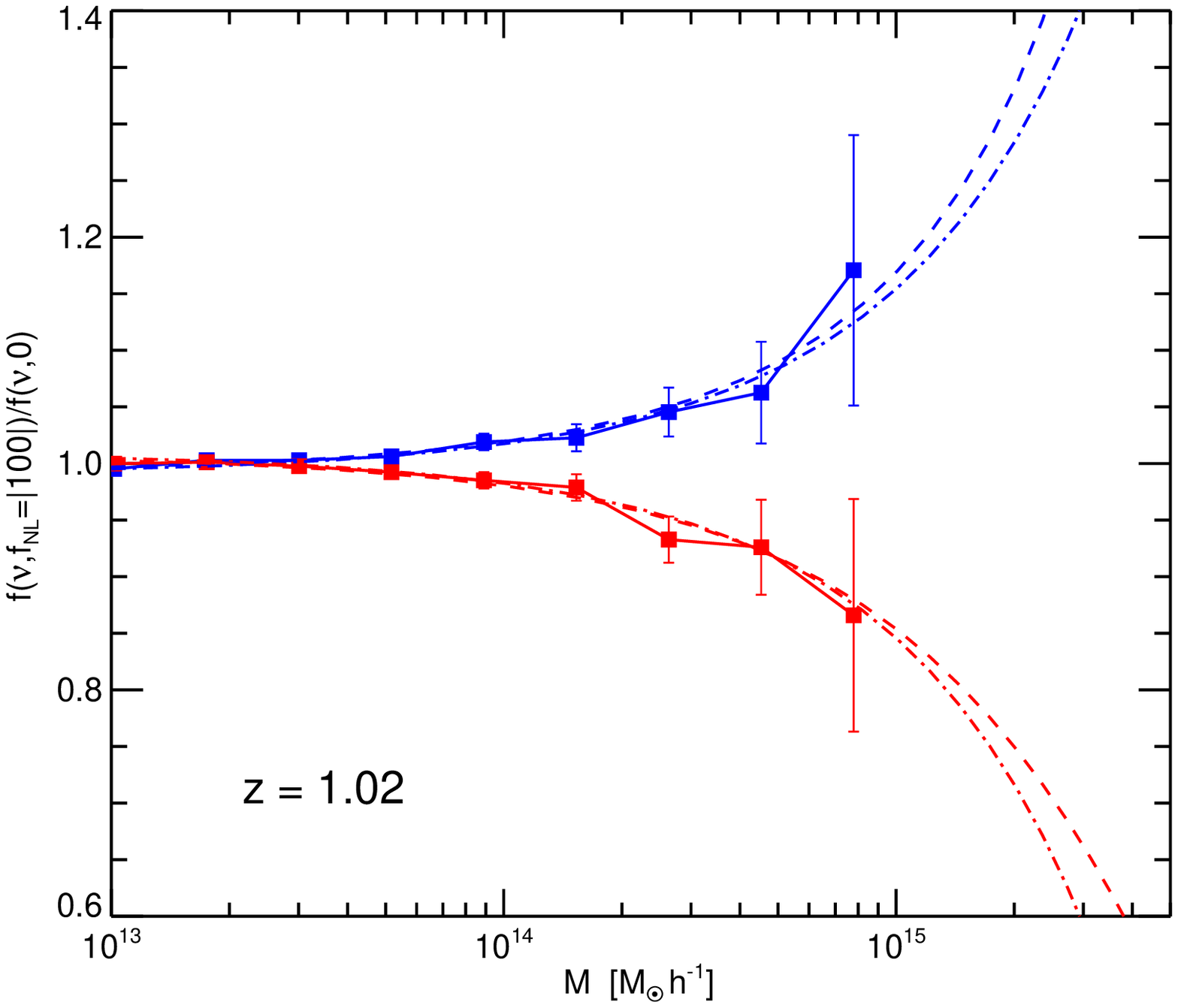}\\
\includegraphics[width=0.45\textwidth]{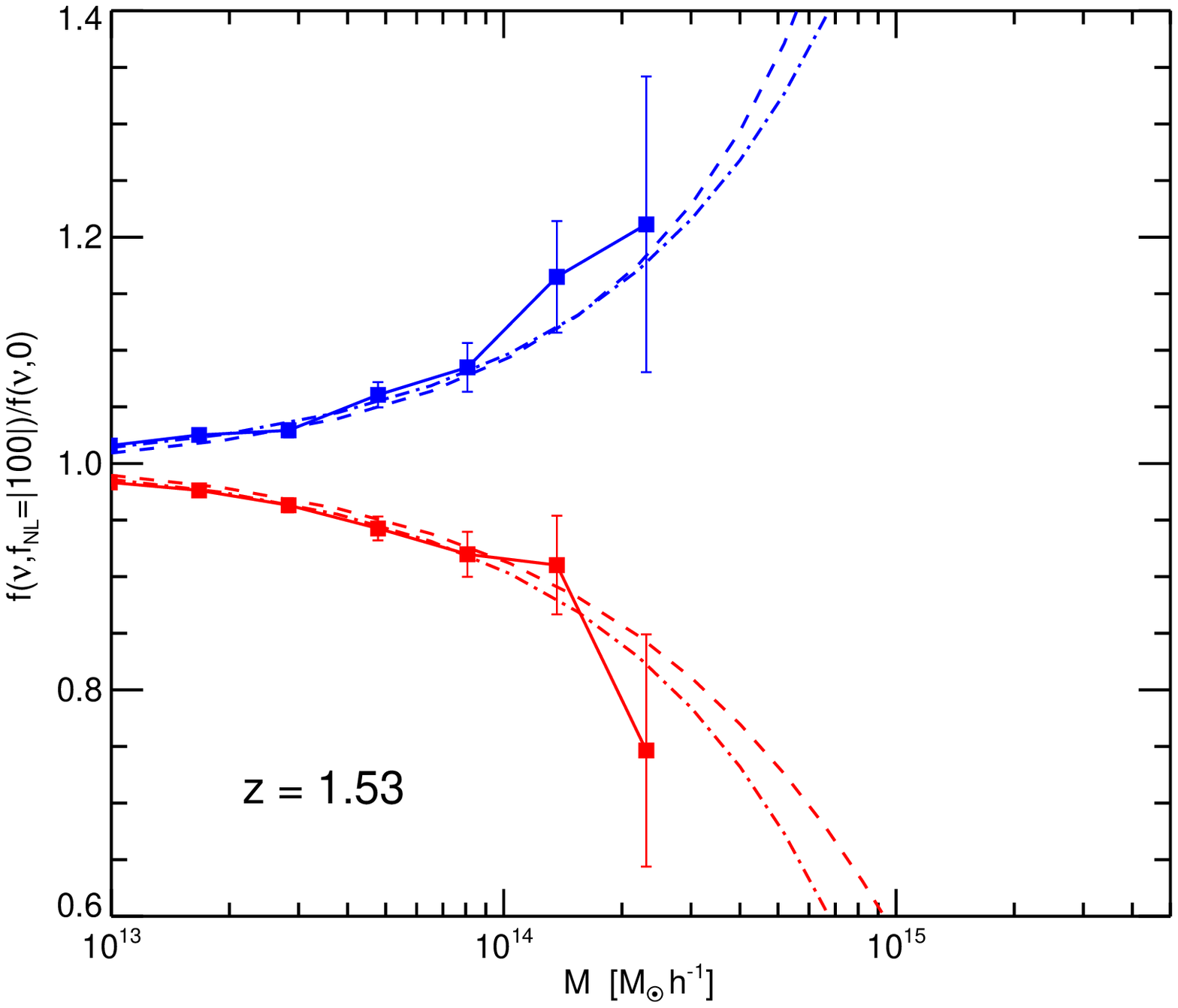}
\includegraphics[width=0.45\textwidth]{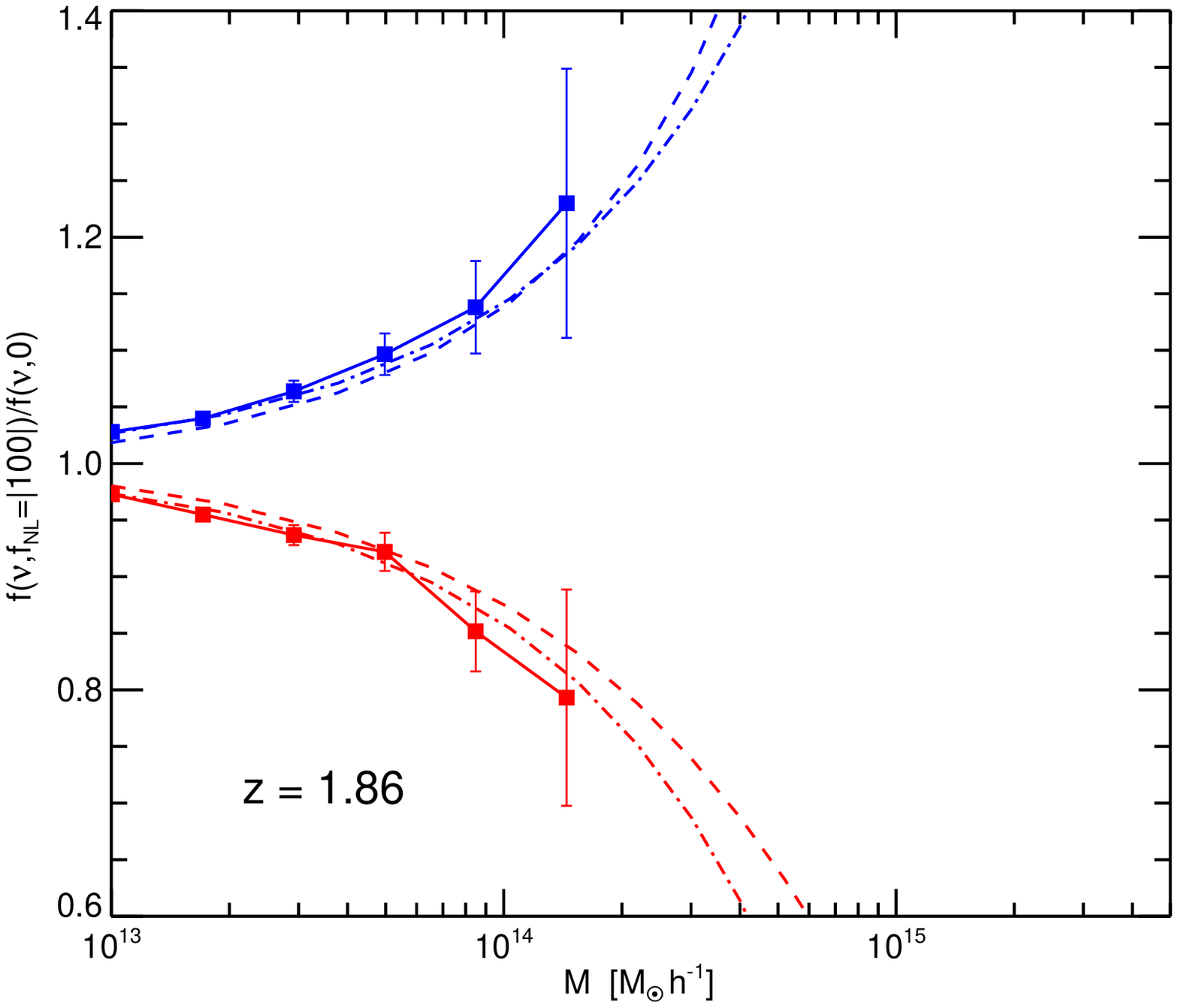}
\end{center}
\caption{Ratio of the non-Gaussian ($f_{NL}=\pm 100$) to Gaussian mass
function for different redshift snapshots: top left $z=0.61$; top right
$z=1.02$; bottom left $z=1.53$; bottom right $z=1.86$. The dashed line
is the mass function of Matarrese, Verde \& Jimenez (2001) and the dot-dashed lines are that
of LoVerde et al. (2008), both including the $q$-correction.}
\label{fig:massfn52}
\end{figure*}

\begin{figure*}
\begin{center}
\includegraphics[width=0.45\textwidth]{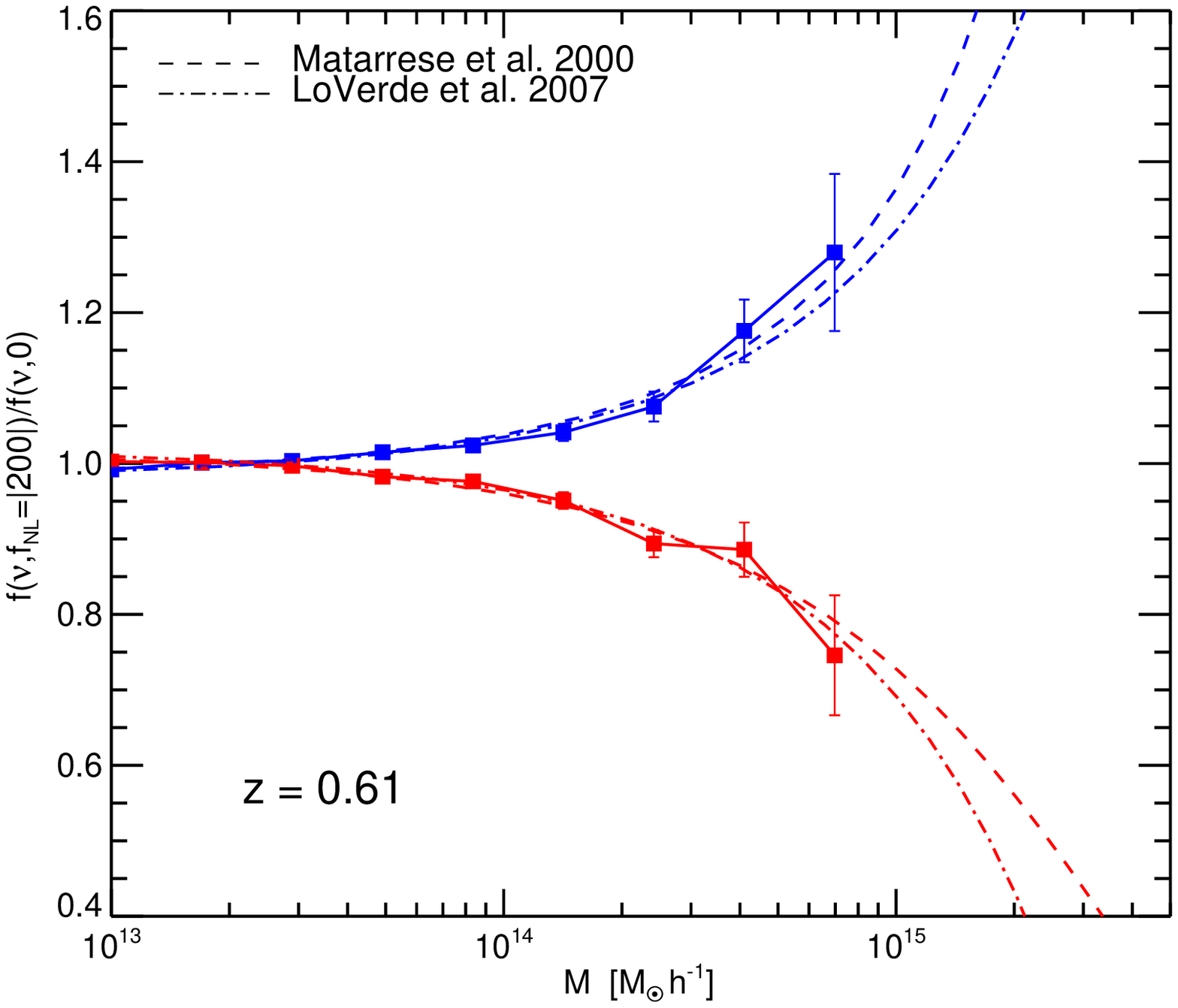}
\includegraphics[width=0.45\textwidth]{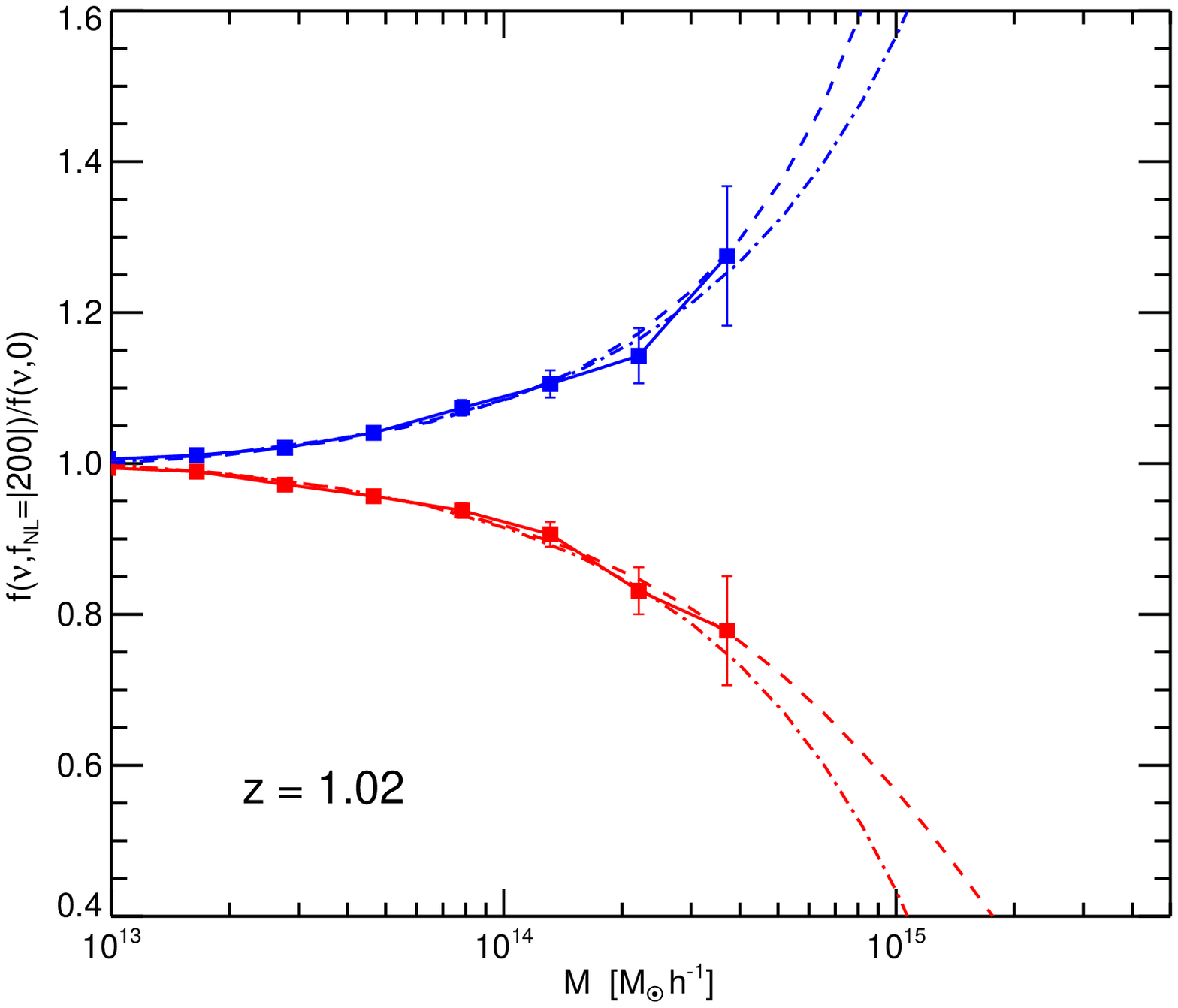}\\
\includegraphics[width=0.45\textwidth]{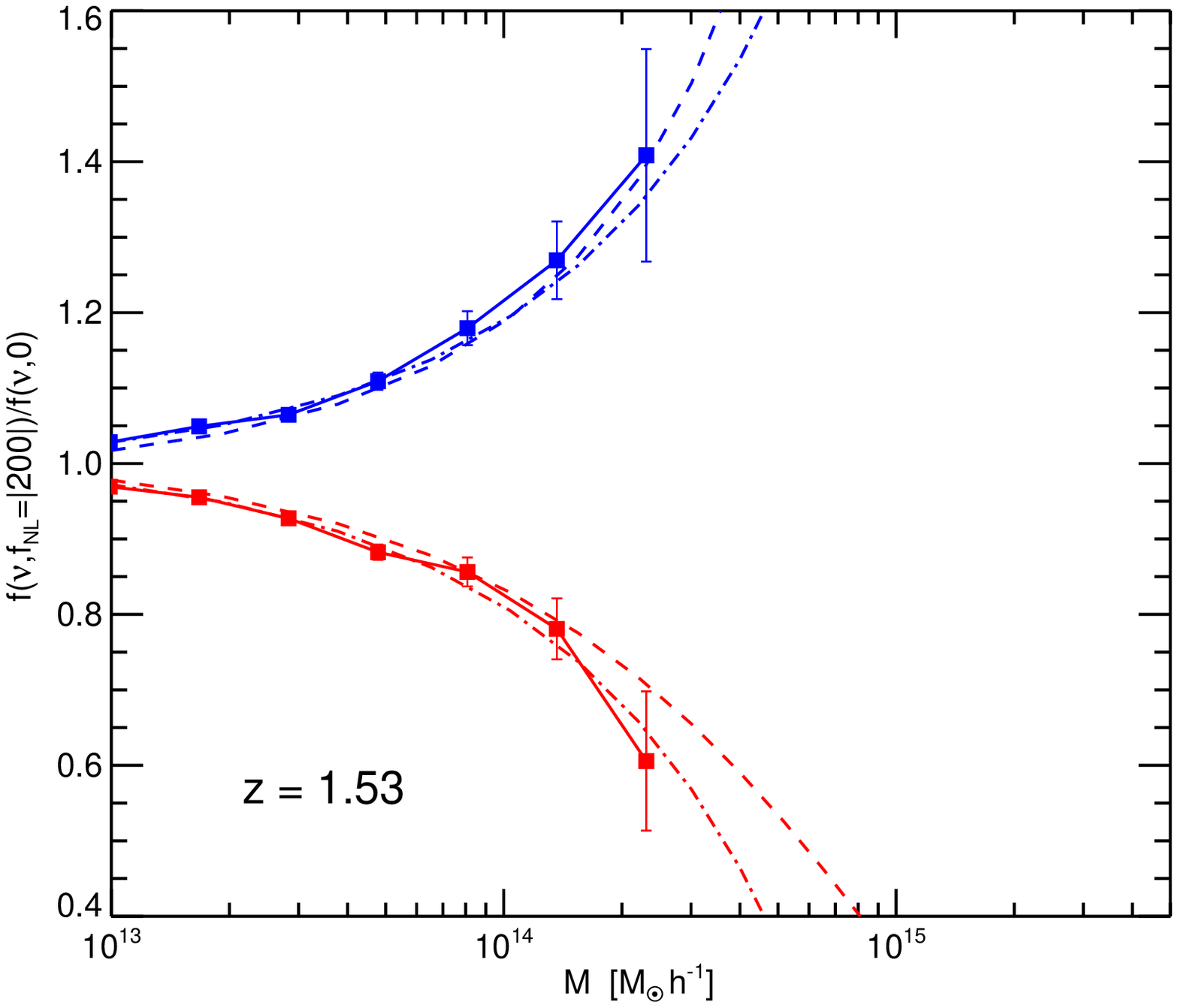}
\includegraphics[width=0.45\textwidth]{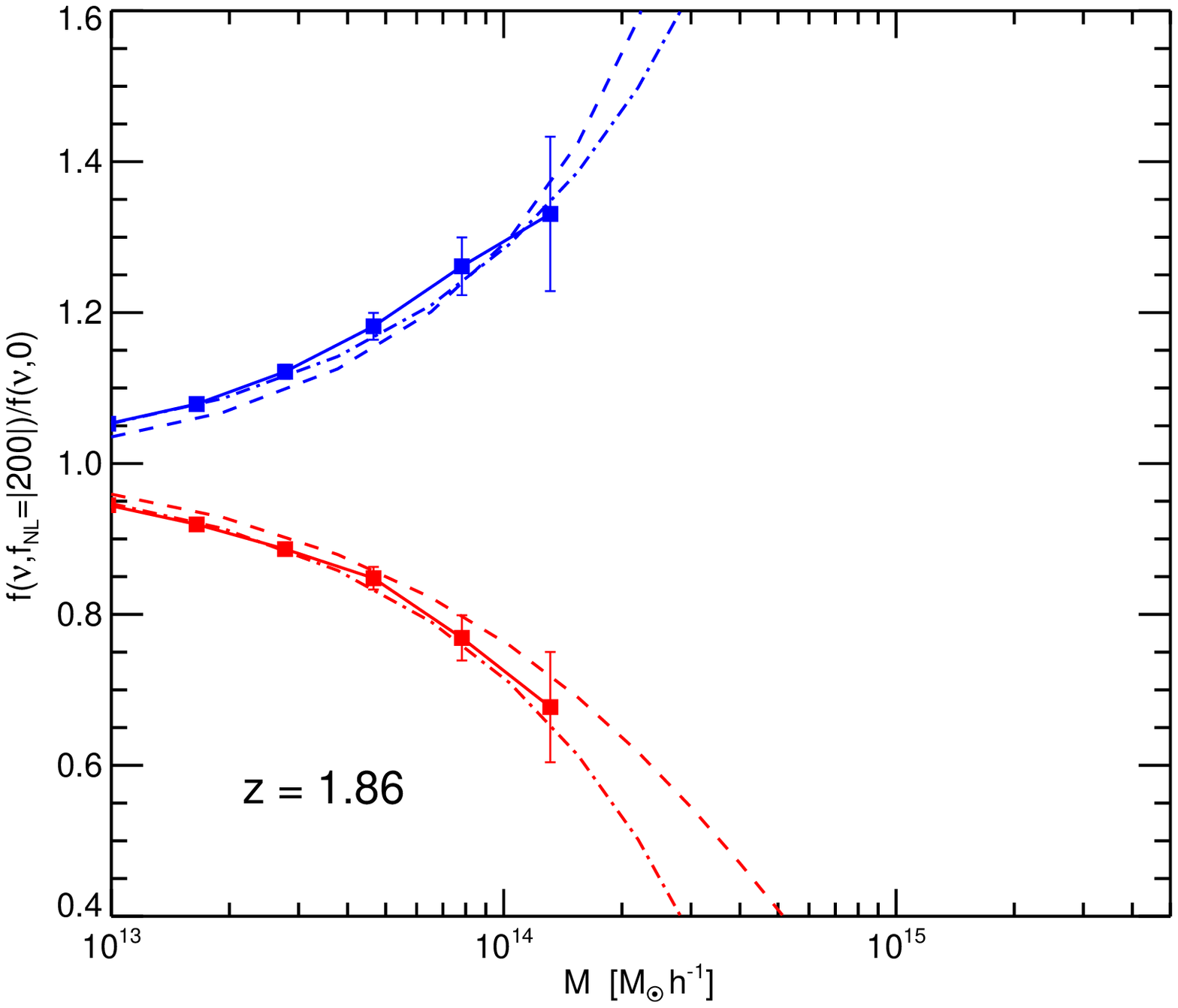}
\end{center}
\caption{Ratio of the non-Gaussian ($f_{NL}=\pm 200$) to Gaussian mass
function for different redshift snapshots: top left $z=0.61$; top right
$z=1.02$; bottom left $z=1.53$; bottom right $z=1.86$. The dashed line
is the mass function of Matarrese, Verde \& Jimenez (2001) and the dot-dashed lines are that of LoVerde et al. (2008), both including our $q$-correction. }
\label{fig:massfn100}
\end{figure*}

\section{Mass function}

We compare the halo mass function of the non-Gaussian simulations with 
the theoretical predictions of Eqs.~(\ref{eq:NGmassfn}), 
(\ref{eq:ratioMVJellips}) and (\ref{eq:ratioLoVellips}) that is, 
including our {\it ansatz} for the the non-spherical collapse correction: 
$\delta_c\longrightarrow \sqrt{q}\delta_c$.
For clarity we show here the non-Gaussian to Gaussian mass function ratio, 
i.e. the factor $R_{NG}(M,z)$.
The comparison between theory and simulations results is shown in 
Fig.~\ref{fig:massfn52} for a few redshift snapshots and for 
$f_{NL}=\pm 100$, and in Fig.~\ref{fig:massfn100} for 
$f_{NL}=\pm 200$ for the same redshifts. 
Dashed lines are the mass function of 
\cite{MVJ00}--Eq.~(\ref{eq:ratioMVJellips}) -- and dot-dashed lines are 
that of \cite{Loverdeetal07}--Eq.~(\ref{eq:ratioLoVellips})--. 

Contrary to \cite{KNS07} and \cite{DDHS07}, we conclude that both 
\cite{MVJ00} and \cite{Loverdeetal07} are good descriptions of the 
non-Gaussian correction to the mass function, once the correction for 
non-spherical collapse is included. 

Figs.~\ref{fig:massfn52} and \ref{fig:massfn100} seem to indicate that 
\cite{Loverdeetal07} may be a better fit for small masses and 
\cite{MVJ00} at high masses. This is not surprising: the Edgeworth expansion  
works well away from the extreme tails of the distribution 
(i.e. for moderate $\delta_c/\sigma_M$), while the saddle-point-approximation 
used in \cite{MVJ00}, is expected to work better at the very tails of the 
distribution (very high $\delta_c/\sigma_M$). We expect that the mass function 
of \cite{MVJ00} will be a better fit at very high masses or larger $f_{NL}$. 
This will be further explored in future work. 

\section{Gaussian halo bias, and the effect of mergers}

\begin{figure}
\includegraphics[width=0.45\textwidth]{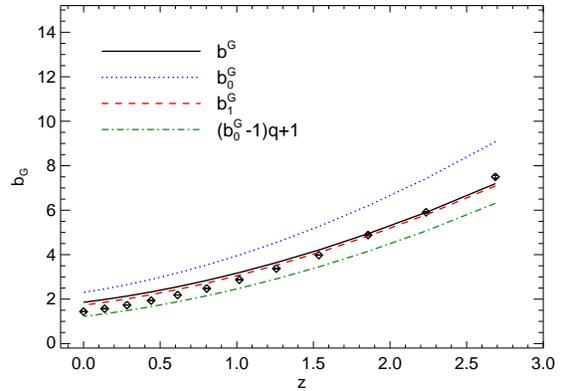}
\caption{Black solid line: the large-scale Gaussian halo Eulerian bias $b^G$ 
of Eq.~(\ref{eq:gaussianbias}). Blue dotted line: the Gaussian halo bias
$b_0^G\equiv 1+\delta_c(z_o)/(\sigma_M^2 D(z_o))$ as defined in Efsthatiou et al. (1988)
and Kaiser (1984). Green dotted-dashed line: $1+q(b_0^G-1)$. Red dashed
line: the contribution $b_1^G\equiv 1+
[q\delta_c(z_f)/\sigma_M^2-1/\delta_c(z_f)]/D(z_o)$ to the total bias of
Eq.~(\ref{eq:gaussianbias}).}
\label{fig:gaussianbias}
\end{figure}

The large-scale, linear halo Eulerian bias  for the Gaussian case is 
\citep{Mo et al. 1997, Scoccimarro et al. 2001}
\begin{equation}
b^G= 1 + \frac{1}{D(z_o)}\left[\frac{q\delta_c(z_f)}{\sigma_M^2}
-\frac{1}{\delta_c(z_f)}\right] +
\label{eq:gaussianbias}
\end{equation}
$$
\frac{2p}{\delta_c(z_f)D(z_o)}\left[1+
\left(\frac{q\delta_c^2(z_f)}{\sigma_M^2}\right)^p\right]^{-1} \;,
$$
where $q=0.75$ and $p=0.3$, account for non-spherical collapse and are a 
fit to numerical simulations.
Here, $\sigma_M$ denotes the {\it rms} value of the dark matter fluctuation 
field smoothed on a scale $R$ corresponding to the Lagrangian radius of 
the halos of mass $M$; $z_f$ denotes the halo formation redshift and 
$z_o$ denotes the halo observation redshift. As we are interested in massive 
halos, we expect that $z_f\simeq z_o$. 
As the non-Gaussian halo bias correction is proportional to $b^G-1$, 
the dependence of $b^G$ on whether the selected halos underwent a recent 
merger (i.e. $z_f \sim z_o$) or are old halos (i.e. $z_f\gg z_o$) affects 
the amplitude of the non-Gaussian correction \citep{slosaretal08, CVM08}.
Before we trust our simulation to accurately describe the non-Gaussian halo 
bias we check whether we recover the Gaussian one and whether the linear halo 
bias approximation is a good description for the scales, redshifts and 
mass-ranges we are interested in. 
\cite{gaospringelwhite} show that analytical predictions for the Gaussian halo 
bias are in reasonable agreement with simulations and that the bias for 
low-mass halos shows strong dependence on formation time but high mass halos 
(the ones we are interested in) do not. 
The halo bias for the Gaussian simulation and the comparison with the theory 
prediction is shown in Fig.~\ref{fig:gaussianbias}. Except for the Gaussian 
halo bias $b_0^G\equiv 1+\delta_c(z_o)/(\sigma_M^2 D(z_o))$ defined in 
\cite{Eetal88} and \cite{Kaiser84} indicated by the dotted (blue) line, the 
simulated data agree with the theoretical expectations at different redshifts. 
In particular, in Fig.~\ref{fig:gaussianbias}, the black solid line represents 
the total Gaussian bias of Eq.~(\ref{eq:gaussianbias}), the dashed (red) line 
represents the contribution from the first line of Eq.~(\ref{eq:gaussianbias}),
and, finally, the dot-dashed (green) line is $1+q(b_0^G-1)$. 
The small difference when using  $z_f \sim z_o$ implies that, for the 
Gaussian halo bias of very massive halos ($M\gs 10^{13}$M$_\odot$), it is 
reliable to assume that the correction from the ``non-spherical 
collapse'' can be encapsulated in the factor $q$ in front of 
$\delta_c(z_o)/(\sigma_M^2 D(z_o))$.

\section{Non-Gaussian halo bias}
\label{sec:ngbias}

Eq.~(\ref{eq:ngcorrection}) shows that the redshift and scale dependence of 
the non-Gaussian correction can be factorized as a term that depends only on 
redshift and one that depends only on $k$ and $M$. The $M$-dependence is 
expected to be  very weak at large scales ($k<0.03$ h/Mpc). Here we will test 
the mass, scale and redshift dependence of the non-Gaussian halo bias and 
we calibrate its normalization on the simulations.
\begin{figure}
\includegraphics[width=0.45\textwidth]{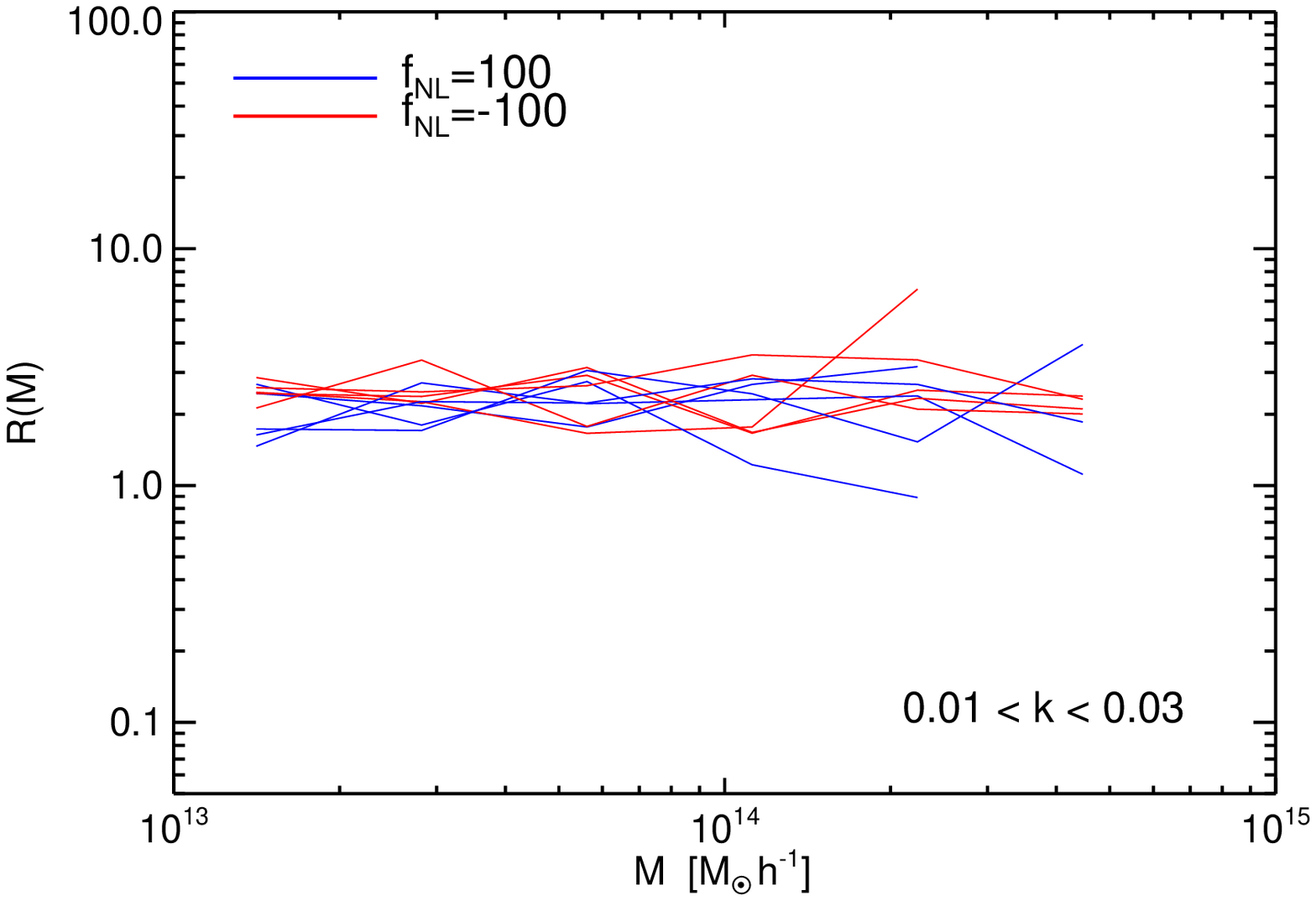}
\caption{Weak mass dependence of $\Delta b /b$ at scales $k<0.03$ h/Mpc. 
Different lines correspond to different redshift snapshots between $z=0$ and 
$z=1.5$. The overall normalization is arbitrary.}
\label{fig:massdepn}
\end{figure}

In Fig.~\ref{fig:massdepn} we show the dependence on halo mass of 
$\Delta b/b_L$. We define the quantity
\begin{equation}
{\cal R}(M)=\left(\frac{\Delta b}{b_L}\right)_{s} 
\left(\frac{\Delta b}{b_L}\right)_{theory}^{-1}\;, 
\end{equation}
where $(\Delta b/b_L)_{theory}$ is given by Eq. (\ref{eq:ngcorrection}). 
To study the mass dependence, we evaluate the theory at fixed mass 
$\widehat{M}=10^{14}M_{\odot}$. We compute the bias from the
simulations taking halos in six different mass bins. Fig.~\ref{fig:massdepn} includes only 
scales $k<0.03$ $h$/Mpc, different lines correspond to different redshift 
snapshots between $z=0$ and $z=1.5$.  As expected, there is no noticeable 
dependence on halo mass.

Having confirmed the expected weak dependence on halo mass for masses 
$M> 10^{13}M_{\odot}/h$ and on scales $k<0.03$ $h$/Mpc, we can study the 
redshift and scale dependence of $\Delta b/b_L$, considering halos of 
different masses above  $10^{13}M_{\odot}/h$.

\begin{figure}
\includegraphics[width=0.45\textwidth]{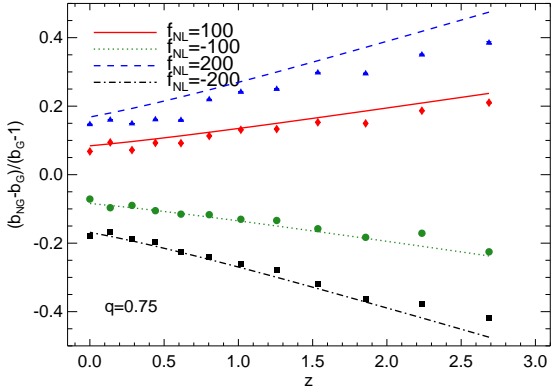}
\caption{The redshift dependence of the non-Gaussian correction to the halo 
bias: points are the values measured from the simulations, lines are the theoretical predictions, Eq.~(\ref{eq:ngcorrection2}). Only $k<0.026$ $h/Mpc$ were used.}
\label{fig:FR}
\end{figure}

The redshift dependence of $\Delta b/b_L$, $(\Delta b/b^G_L)_{s} [2 f_{NL} 
\alpha_M(k) q]^{-1}$ is shown in Fig.~\ref{fig:FR} where 
$M> 10^{13}M_{\odot}/h$ and scales $k< 0.026$ $h$/Mpc were used. 
In applying the  correction $\delta_c/\sigma_M^2
\longrightarrow q \delta_c/\sigma_M^2$  to $\Delta b/(b_G-1)$ we have  actually corrected $b^G_0$, i.e. we have  employed the same approximation used for the green dot-dashed line of Fig. 8, giving Eq.~(\ref{eq:ngcorrection2}). Eq.(9) in fact is only the consequence of our correction to the Gaussian halo  bias. Note that the approximation  $z_f \sim z_0$ we employed here is expected to hold for rare--massive--halos  and 
%(which correspond to the same approximation as for the green dot-dashed line in Fig. 8).
Fig. 8 shows that this is a good approximation. A detailed study of the 
dependence of the non-Gaussian halo bias correction on the formation 
redshift of the halos will be presented elsewhere.

There seems to be an indication that the $q$-correction 
factor for the large-scale bias correction may slightly depend on the value 
of $f_{NL}$: in particular the figure shows that it could be slightly smaller   
than $q$ for $f_{NL}$ large and negative and smaller for $f_{NL}$ 
large and positive. This is not unexpected: the presence of non-Gaussianity 
may alter the dynamics of non-spherical collapse (e.g., through tidal forces 
-- see e.g., \cite{Desjaquesb} -- or by significantly changing the redshift 
for collapse with respect to the Gaussian case). At this stage, however, 
this trend is not highly significant and further study will be left to 
future work.

\begin{figure}
\includegraphics[width=0.45\textwidth]{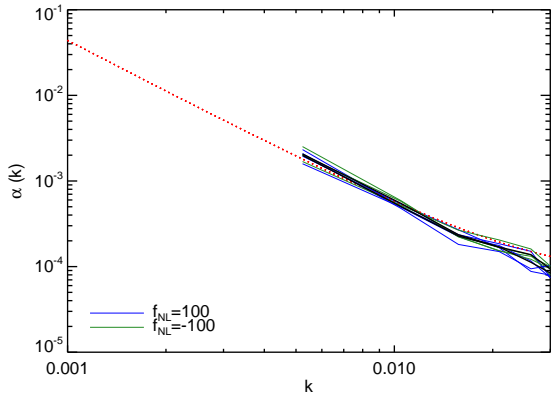}
 \caption{Scale dependence of Eq.~(\ref{eq:ngcorrection2}). The thin lines 
correspond to different redshifts for halos with mass above $10^{13} M_\odot/h$ 
and the thick black line  is their average. The dotted line is the theory prediction with $q=0.75$. 
At $k>0.03$h/Mpc the effect of non-Gaussianity is very small and the 
measurement become extremely noisy.}
 \label{fig:alphak}
 \end{figure}
Finally, we show the scale dependence of Eq.~(\ref{eq:ngcorrection2}),  
$(\Delta b/b^G_L)_{s} [2 f_{NL} \delta_c(z) q]^{-1}$, in Fig.~\ref{fig:alphak}.
The thin lines correspond to different redshifts and the thick  black
line to their average. The dotted line is the theory prediction with $q=0.75$.
Note that  there is an excellent agreement on the scales of interest, 
e.g., $k<0.03$ $h$/Mpc. On smaller scales the effect of non-Gaussianity is very small and the measurement  become extremely noisy. These results are in qualitative agreement with
the findings of \citet{Pillepich}.

We conclude that Eq.~(\ref{eq:ngcorrection2}), with $q \sim 0.75$, 
provides a good fit to non-Gaussian simulations.

\begin{figure*}
\begin{center}
\includegraphics[width=0.45\textwidth]{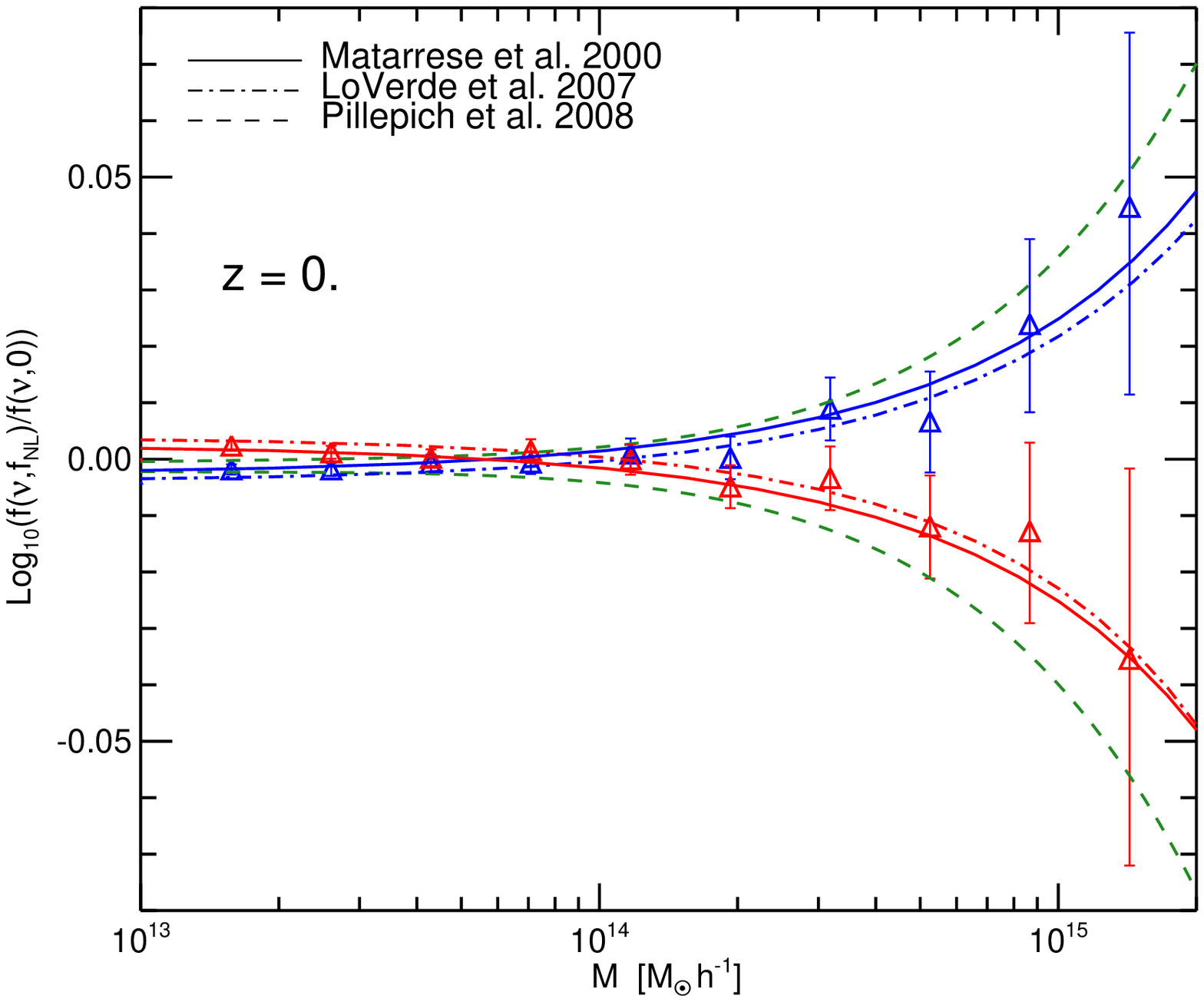}
\includegraphics[width=0.45\textwidth]{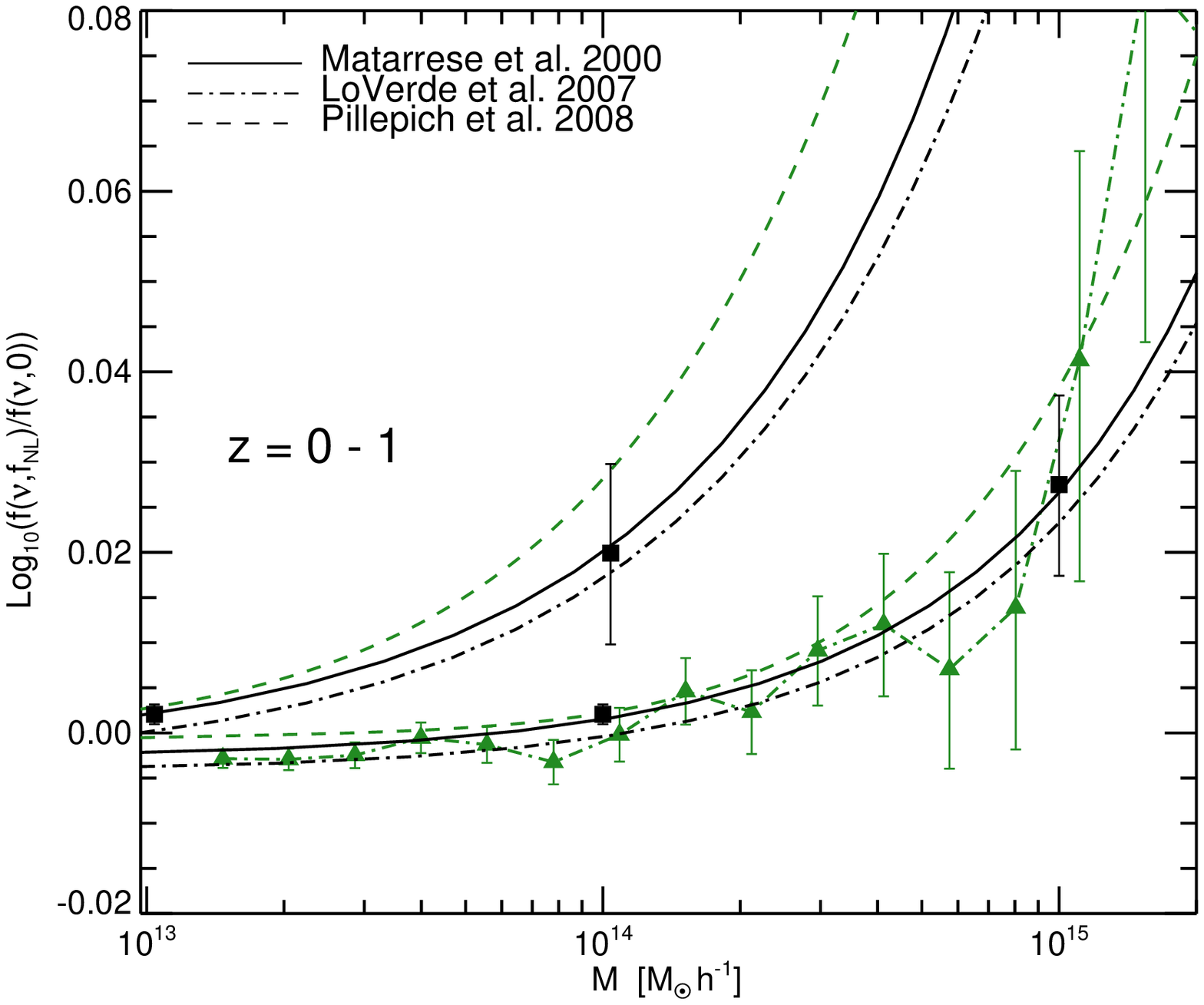}
\end{center}
\caption{Left panel: comparison between simulations points and fits from this 
work and the polynomial fit of \citet{Pillepich}. Right panel: 
Our correction to the Matarrese, Verde \& Jimenez (2001) and LoVerde et al (2008) non-Gaussian 
mass function fits, the polynomial fit of Pillepich, Porciani \& Zahn (2009) and points from 
Desjaques et al (2009) (black squares) and Pillepich, Porciani \& Zahn (2009) (green triangles).  See text for more details. }
\label{fig:comparisonth-sim}
\end{figure*}

\section{Comparison with previous work}

After the discussion of \S 2, it should be clear that if $f_{NL}^{CMB}$ were 
used in the theoretical predictions or $S_3$ were {\it not} linearly 
extrapolated to $z=0$, then any constraints on non-Gaussianity so obtained 
would have to be re-scaled by a factor $\sim 1.3$. This seems to be the case 
of some work in the literature.
On the other hand the  q-correction factor effectively 
introduces a re-scaling of a factor $\sim 0.75^{3/2}=0.65$ for the mass function 
case and $0.75$ for the bias case.  
It is a coincidence that, for the halo bias, $1.3\times 0.75 \sim 1$, thus the 
$f_{NL}$ normalization mistake cancels out with the spherical collapse 
approximation error. This fortuitous cancellation does not happen to the same 
level in the mass function $0.75^{3/2}\times 1.3\sim 0.8$, explaining perhaps 
some of the claimed discrepancy of the simulations with the analytic mass 
function predictions, and the claimed agreement with the halo bias 
predictions. Another possible source of inaccuracy would be an inconsistent treatment of the redshft evolution of $\delta_c$ and $S_3$ (see discussion is \S 2).

 In Fig.~\ref{fig:comparisonth-sim} we compare our theoretical 
predictions with the results presented in \cite{Pillepich} and \cite{Desjaques}.
The left panel shows our simulation results at $z=0$ for $f_{NL} = \pm 100$ 
and our theoretical predictions. Additionally we show the fit presented by
\cite{Pillepich}, Eqs. (8) and (9), evaluated for the suitable values of $f_{NL}$ 
accounting for the different notations for $f_{NL}$. We also adopt our
cosmological parameters when converting $\sigma_M$ to $M$. The right panel
shows the simulation results presented in \cite{Pillepich} at z=0 and
their fitting formula at $z=0$ and $z=1$. We over plot
our theoretical models evaluated for their cosmological parameters and
for the corresponding values of $f_{NL}$. Moreover we add the data points 
from \cite{Desjaques} for $z=0$ and $z=1$, suitably rescaled by the differences of
the $f_{NL}$ value used. 
The mass function  fits of  Fig. 12 differ for large masses, in the regime where simulations errors become large;  the fits are however  consistent given the individual points error-bars.

Our theoretical formulae for the non-Gaussian mass function (Eqs. 5, 6, 7) and for 
the non-Gaussian halo bias (Eq. 9) are physically motivated expressions that 
have been tested on N-body simulations. They have the advantage over fitting formulae 
that they can be more robustly interpolated and extrapolated to cosmologies and 
parameters that have not been directly simulated and they are more robust 
over parameters ranges where the simulations have low signal-to-noise.
Compared to simple fitting formulae, Eq. 6,7 and 9  have the disadvantage 
that they require the calculation of some numerical integrals. To overcome this, 
we supply tabulated values for $S_{3,M}^{(1)}$, $\sigma_M$ and $\alpha_M$ for a 
WMAP5 cosmology in the range of interest at  www.ice.csic.es/personal/verde/nongaussian.html. 

The $q$-correction  we find here has implications for 
previously reported and forecasted constraints on non-Gaussianity. 
In Table 1 we report  present and forecasted constraints on $f_{\rm NL}$ 
from the literature rescaled to $f_{\rm NL}^{CMB}$ and corrected for 
our factor $q$. 

This confirms that constraints on $f_{NL}$ achievable using the 
non-Gaussian halo bias are competitive with CMB constraints ($f_{NL}\sim 5$ 
for Planck and $f_{NL} \sim 3$ for a CMBPol-type mission, \cite{babich2004,Yadav2007}).
 
\begin{table}%[H] add [H] placement to break table across pages
 \caption{\label{tab:currentlocalfnl} Current and forecasted constraints 
on $f_{\rm NL}^{CMB}$ }
% \begin{ruledtabular}
 \begin{tabular}{ccc}
 &Measurements&\\
 Data/method & $f_{\rm NL}, 1-2\sigma$ errors & reference\\
 \hline
 \hline
 &&\\
 Photo LRG--bias & $84^{+54+85}_{-101-331}$ & Slosar et al. 2008 \\
 & &\\
 Spectro LRG--bias & $93^{+74 +139}_{-83 -191}$& Slosar et al. 2008\\
 & &\\
 QSO - bias& $11^{+26+47}_{-37-77}$& Slosar et al. 2008\\
 &&\\
 combined & $37^{+23+42}_{-26-57}$&Slosar et al. 2008\\
 &&\\
 \hline
 &&\\
 NVSS--ISW  &$140^{+647+755}_{-337-1157}$ &  Slosar et al. 2008\\
 &&\\
 NVSS--ISW & $ 272\pm 127$ (2-$\sigma$)&$\!\!\!\!$Afshordi\&Tolley 2008\\
 &&\\
 \hline
& Forecasts&\\
Data/method & $\Delta f_{\rm NL}(1-\sigma)$ & reference\\
 \hline
 &&\\
 BOSS--bias & $18$& Carbone et al 2008\\
 &&\\
 ADEPT/Euclid--bias &$1.5$ &  Carbone et al. 2008 \\
 &&\\
 PANNStarrs--bias  &$3.5$ & Carbone et al. 2008\\
 &&\\
 LSST--bias  &$0.7$ & Carbone et al. 2008\\
&&\\
\hline
&&\\
 LSST-ISW&$10.$&$\!\!\!\!$ Afshordi\&Tolley 2008\\
 &&\\
 \end{tabular}
 %\end{ruledtabular}
 \end{table}

\section{Conclusions}
We have considered $(1.2$ Gpc/h)$^3$ size and $960^3$ particles N-body 
simulations with non-Gaussian initial conditions, with non-Gaussianity 
parameter $f_{NL}=\pm 100$, $f_{NL}=\pm 200$ and a reference Gaussian 
simulation ($f_{NL}=0$). The clustering properties and the abundance of 
the simulation's halos were then compared with independent simulations 
and theoretical predictions.
We find good agreement between different simulations, indicating that 
the initial conditions set-up is under control. 
%Contrary to \cite{KNS07, DDHS07}, but in broad agreement with 
%\cite{Grossietal07}
We find that the Press-Schechter-based 
description of the non-Gaussian correction to the Gaussian mass function 
of \cite{MVJ00} and \cite{Loverdeetal07} is a good fit to the simulations, 
provided that:\\
{\it a)} The Press-Schechter-based description is used to compute the ratio 
between  Gaussian and non-Gaussian mass function.\\
{\it b)} The critical density $\delta_c$ is corrected to account for 
non-spherical  collapse dynamics. \\
This is summarized in our Eq.~(\ref{eq:NGmassfn}) and in 
Eqs.~(\ref{eq:ratioMVJellips}) and (\ref{eq:ratioLoVellips}) for 
the non-Gaussian mass functions of \cite{MVJ00} and \cite{Loverdeetal07}, 
respectively.
For large thresholds this correction is equivalent to a re-scaling of 
the spherical collapse threshold  $\delta_c\times \sqrt{q}$ where $q=0.75$.
The $q$-correction is thus equivalent to a reduction of 
$f_{NL}$ by a factor $\sim 1.5$ because in the mass function, to leading 
order $f_{NL}$ multiplies $\delta_c^3$.

We find that the non-Gaussian halo bias prescription of \citet{DDHS07}, 
\citet{MV08}, \citet{slosaretal08} and \cite{AfshordiTolley08} 
provides a good description of the scaling of the large-scale halo 
clustering of the simulations. In particular, we have tested separately 
the predicted redshift, scale and $f_{NL}$ dependence. 
The overall amplitude of the effect, however, should be corrected by a 
factor $\sim q$ which can also be understood in the context of ellipsoidal 
collapse or as a modification to the excursion set ansatz and the sharp-k space   filtering  (see Eq.~(\ref{eq:ngcorrection2})). 
There is an indication that this correction may be slightly dependent on 
$f_{\rm NL}$. This is not unexpected, but the signal-to-noise of the effect 
is too small in the current simulations to draw robust conclusions.
We also find that on large ($k<0.03$ $h$/Mpc) scales, as expected, 
the fractional correction to the non-Gaussian halo bias is independent 
of mass. On smaller scales a dependence on mass is expected, but 
the simulations do not have sufficient signal-to-noise to verify it.
The $q$--correction to the non-Gaussian halo bias modifies 
current and forecasted constraints reported in the literature as 
indicated in our Table 1.

The formulae we presented here for the non-Gaussian  mass function  (Eq. 5, 6,7) 
and non-Gaussian halo bias (Eq. 9) are physically motivated expressions which 
provide good fits to a suite of N-body simulations. As such, they can be more 
robustly interpolated and extrapolated than simple fitting functions 
(in www.ice.csic.es/personal/verde/nongaussian.html we provide useful quantities for ease of use of these equations). 
We confirm that the non-Gaussian halo bias offers a robust and highly 
competitive test of primordial non-Gaussianity.

\section*{Acknowledgments}
Computations have been performed on the IBM-SP5 at CINECA (Consorzio
Interuniversitario del Nord-Est per il Calcolo Automatico), Bologna,
with CPU time assigned under an INAF-CINECA grant, and on the IBM-SP4
machine at the ``Rechenzentrum der Max-Planck-Gesellschaft'' at the
Max-Planck Institut fuer Plasmaphysik with CPU time assigned to the
``Max-Planck-Institut f\"ur Astrophysik'' and at
the ``Leibniz-Rechenzentrum'' with CPU time assinged to the Project ``h0073''.
The authors thank A. Pillepich and C. Porciani for dicussions and for 
making available the simulations outputs for comparison. 
We also thank N. Dalal, U. Seljak, A. Riotto and M. Maggiore for discussions.
LV is supported by FP7-PEOPLE-2007-4-3-IRG n. 202182 and CSIC I3 grant 
n. 200750I034.
CC is supported through a Beatriu de Pinos grant. This research was 
supported by the DFG cluster of excellence Origin and Structure of 
the Universe.
SM and LM acknowledge partial support by ASI contract I/016/07/0 ``COFIS", 
ASI-INAF I/023/05/0, ASI-INAF I/088/06/0 and ASI contract Planck LFI 
Activity of Phase E2.
LV thanks the Santa Fe cosmology workshop 2008 , LV and CC thank the Galileo 
Galilei Institute for theoretical physics in Florence, where part of this work 
was carried out, and INFN for partial support. 
LV thanks B. Wandelt for discussions.

\bibliographystyle{mn2e}
%\bibliographystyle{apj}
%\bibliography{lvbib}

\end{document}